\documentclass[12pt,preprint]{aastex}

\usepackage{graphicx,natbib,textcomp,caption,subcaption}         
\usepackage[utf8]{inputenc}
\usepackage{xcolor,amsmath}

\slugcomment{ApJ manuscript, \today}

\shorttitle{Multi-stranded loops with transverse waves}
\shortauthors{Magyar, \& Van Doorsselaere}

\begin{document}

\title{The instability and non-existence of multi-stranded loops, when driven by transverse waves}

\author{N. Magyar\thanks{FWO (Fonds Wetenschappelijk Onderzoek) PhD fellow}, T. 
Van Doorsselaere}
\affil{Centre for mathematical Plasma Astrophysics (CmPA), KU 
Leuven, Celestijnenlaan 
200B bus 2400, 3001 Leuven, Belgium; norbert.magyar@wis.kuleuven.be}

\begin{abstract}
In recent years, omni-present transverse waves have been observed in all layers of the solar atmosphere. Coronal loops are often modeled as a collection of individual strands, in order to explain their thermal behaviour and appearance. We perform 3D ideal MHD simulations to study the effect of a continuous small amplitude transverse footpoint driving on the internal structure of a coronal loop composed of strands. The output is also converted to synthetic images, corresponding to the AIA 171\AA\ and 193\AA\ passbands, using \texttt{FoMo}. We show that the multi-stranded loop ceases to exist in the traditional sense of the word, because the plasma is efficiently mixed perpendicularly to the magnetic field, with the Kelvin-Helmholtz instability acting as the main mechanism. The final product of our simulation is mixed loop with density structures on a large range of scales, resembling a power-law. Thus, multi-stranded loops are unstable to driving by transverse waves, and this raises a strong doubt on the usability and applicability of coronal loop models consisting of independent strands.
\end{abstract}


\section{Introduction}
The observations of our most recent imaging satellites show plenty of evidence for complex and fine structure in the solar corona. It is commonly believed that the highly-structured nature of the corona is due to the all-permeating magnetic field. Coronal loops, the building blocks of active regions, probably outline the magnetic field. Even with the advent of imaging instruments with increasingly higher spatial resolution, the question still remains whether the coronal loops we see are fully resolved, i.e. whether they have sub-resolution thermal and spatial structure \citep[just to name a few]{1990Natur.344..842G,1993ApJ...405..767G,1994ApJ...422..381C,
2000SoPh..193...53K,2000ApJ...541.1059A,2007Sci...318.1582R,2012ApJ...745..152A,2013A&A...556A.104P,2014ApJ...797...36S}. 
Studies with models of coronal loops including this sub-structuring, i.e. consisting of a bundle of thin strands, have had success in explaining some of their observed properties, which cannot be explained by monolithic loop models, e.g. a broad DEM \citep{2012ApJ...755L..33B}, uniform filter ratio distribution along warm loops \citep{1999ApJ...517L.155L,2000ApJ...528L..45R}, delay between the appearance of the loop in different filters \citep{2003ApJ...593.1174W, 2012ApJ...753...35V}, the ``fuzzy" appearance of loops in harder energy bands, i.e. in X-rays \citep{2009ApJ...694.1256T}, their apparent constant cross-section \citep{2012A&A...548A...1P}. However, there are still some discrepancies between what these models predict and observed features \citep[see][sect. 4.2 for more discussions]{2010LRSP....7....5R}. In a recent study analyzing Hi-C data \citep{2013Natur.493..501C}, it was concluded that the individual strands in the observed loop should be smaller than 15 km in diameter if they exist, without excluding the possibility that the loops are monolithic structures after all  \citep{2013A&A...556A.104P}. In the multi-stranded loop models, the strands are assumed to have an independent hydrodynamic evolution \citep[e.g.][]{2014arXiv1405.3450R,2015ApJ...811..129B}, and on each of these strands the thermal processes are modelled in great detail. The assumption of independent evolution seems reasonable since the cross-field transport is greatly inhibited in the corona \citep[even though cross-field diffusion can be enhanced,][explaining the typical widths of loops or strands]{2006ApJ...646..615G}. \par
Recently, ubiquitous transverse waves were observed in the chromosphere and corona \citep{2007Sci...317.1192T,2007Sci...318.1574D}. These are interpreted as kink modes \citep{2008ApJ...676L..73V}, with a largely Alfv\'enic character \citep{2009A&A...503..213G, 2012ApJ...753..111G}. These waves have relatively low amplitudes (few to few tens of km/s) and low frequency (50-500 s), showing differences in these values whether they are measured in coronal holes, the quiet Sun, or active regions, the latter generally having shorter periods and lower amplitudes \citep{2011Natur.475..477M,2013A&A...553L..10M,2014ApJ...790L...2T,2015NatCo...6E7813M}. These low-amplitude waves are observed to have little-to-no damping \citep{2012ApJ...751L..27W, 2013A&A...552A..57N}, and this probably means that there is an steady-state in which the dissipated energy balances the energy input of the driver. Although the waves in coronal loops have been modeled as propagating waves \citep{2010ApJ...718L.102V,2010A&A...524A..23T}, there are some observational indications that they are actually standing waves \citep{2013A&A...560A.107A}. \par
It was suggested long before the discovery of ubiquitous transverse waves in the corona that coronal loops must be in a permanent state of Kelvin-Helmholtz instability (KHI) \citep{1983A&A...117..220H}. Resonance layers in coronal loops \citep{1992SoPh..138..233G,2002A&A...394L..39G,2002ApJ...577..475R} present a shear flow velocity profile, which is potentially susceptible to the instability \citep{1991SoPh..134..111U,1994ApJ...421..372K,1994GeoRL..21.2259O,1997SoPh..172...45P}, leading to turbulent broadening of shear layers, small-scale structures, and enhanced dissipation, confirmed by recent full 3D MHD numerical simulations \citep{2008ApJ...687L.115T,2014ApJ...787L..22A,2015ApJ...809...72A,2015A&A...582A.117M}, even for small amplitude transverse oscillations, on the order of the ubiquitous transvere wave amplitudes. Direct observational evidence of the KHI in coronal loops is missing. So far, the instability has been observed in coronal mass ejections and quiescent prominences \citep{2010ApJ...716.1288B,2010SoPh..267...75R,2011ApJ...729L...8F,2011ApJ...734L..11O}. However, \citet{2014ApJ...787L..22A} suggested that the roll-ups and vortices of the KHI around coronal loops could be seen as strands in EUV, thus explaining their apparent stranded nature.  
Another feature of the KHI is the perpendicular mixing of material. The observational evidence suggesting gradual interaction and mixing of plasma at different temperatures in near the apex of coronal loops \citep{2004ApJ...617L..81S,2006SoPh..236..245S,2013ApJ...765L..46K} could be related to this mixing. 

The recent observations of ubiquitous transverse waves and the frequently used multi-strand interpretation of multi-thermal loops raises the question on the effect of multi-stranded loops on transverse waves and vice versa. This has been studied in two-loop systems by \citet{2008ApJ...676..717L,2008A&A...485..849V,2011A&A...525A...4R,2014A&A...562A..38G} and in more complex configurations \citep{2009ApJ...692.1582L,2009ApJ...694..502O,2010ApJ...716.1371L,
2011ApJ...731...73P,2012ApJ...746...31D,2014ApJ...795...18V,2015A&A...582A.120S}. In this paper we investigate the effects that a continuously driven small amplitude transverse oscillation has on the internal structure of a coronal loop composed of smaller strands, allowing for non-linear development of the perturbations and strands.

\section{Numerical Model}

\begin{figure}
    \centering
    	\includegraphics[width=0.5\textwidth]{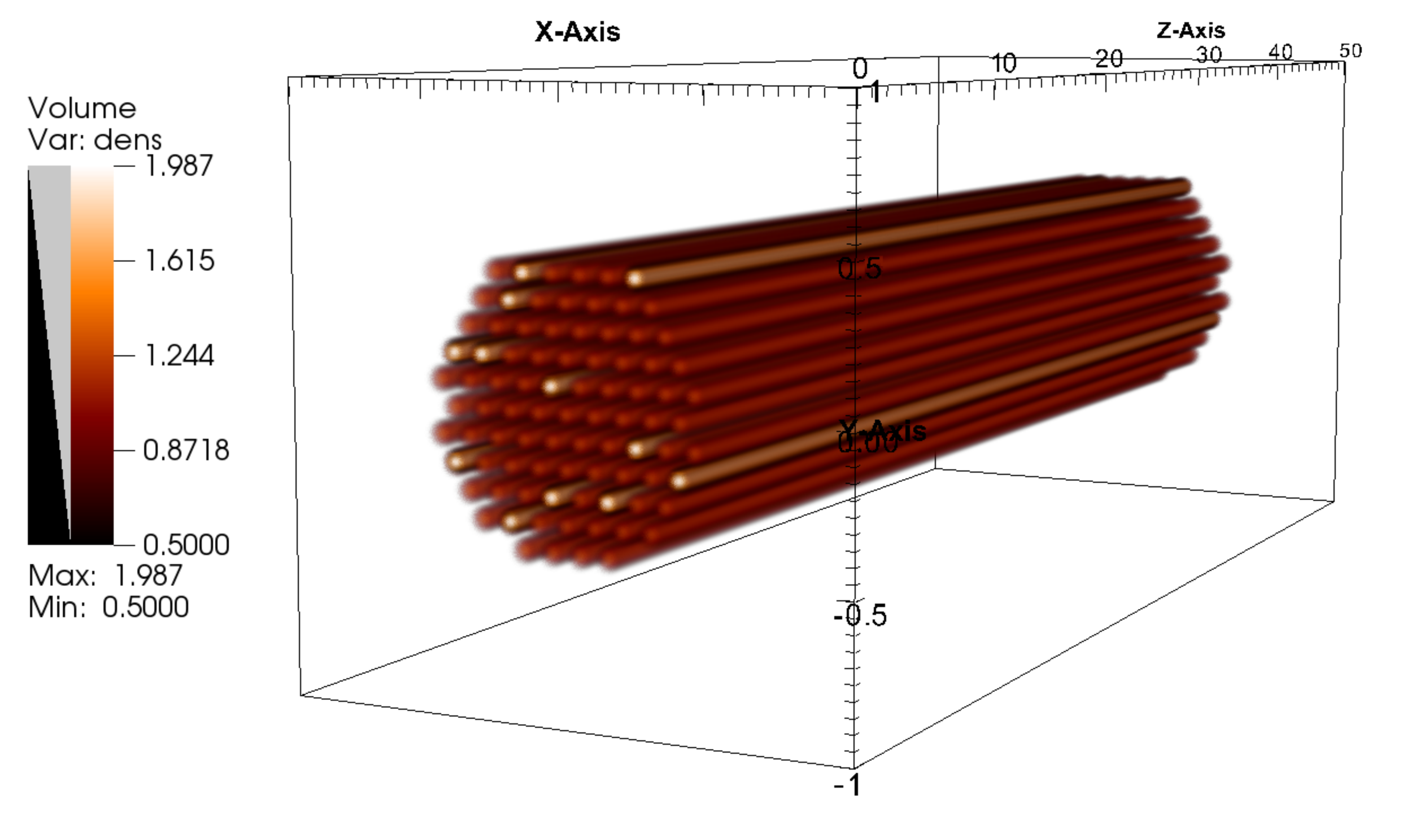}  
    	\caption{Figure showing the numerical box and the initial density by volume ray tracing. Density is in $10^{-12}$ kg m$^{-3}$, while axis units in Mm. Note the high aspect ratio of the axes. The grey-black bar next to the color bar indicates the transparency of plasma varying with density: black denotes transparent, grey fully opaque.}
    	\label{img1}
 \end{figure}

Our model consists of a bundle of straight, hexagonally closely packed thin strands (cylinders) of enhanced density plasma, and a less dense background plasma (see Fig.~\ref{img1}). It is based on the honeycomb model, introduced in \citet{2013A&A...556A.104P}. The whole domain is permeated by a homogeneous and straight magnetic field, parallel to the strands, in the $z$-direction. 
We thus consider a ``macro-loop'' (with a radius $R$, and centered on the axis $x=y=0$). The space in this macro-loop (or loop for short) is filled with strands in the honeycomb structure. The centre of each strand is denoted as $(x=x_\mathrm{s},y=y_\mathrm{s})$, and all strands have a radius $R_\mathrm{s}$. Then, the continuously varying density in each strand is defined as 
\begin{multline}
  \rho = \rho_\mathrm{e}+(\rho_\mathrm{i}-\rho_\mathrm{e})\ \mathrm{cos}\left(\frac{\pi \sqrt{(x-x_\mathrm{s})^2+(y-y_\mathrm{s})^2}}{2 R_\mathrm{s}}\right),\\ \mbox{for } (x-x_\mathrm{s})^2+(y-y_\mathrm{s})^2 \leq R_\mathrm{s}^2
\end{multline}
where $\rho_\mathrm{e}, \rho_\mathrm{i}$ are the background and peak densities, respectively. Some of the strands, randomly chosen with a probability $p = 0.2$, are nearly twice as dense. The thermal pressure is constant throughout the domain (gravity is neglected), implying that less dense plasma is hotter. The distance between the center of any two neighboring strands is $2R_\mathrm{s}$, thus the strands are tightly packed.
The values of the principal physical parameters used in the setup are listed in Table~\ref{table1}.

  \begin{table}
  \caption{\label{table1}The values of principal physical parameters used in the simulations.}
  	\centering
  	\begin{tabular}{lccc}
  		\hline\hline
  		Parameter & First setup\\ 
  		\hline
  		Loop length($L$)& 50 Mm\\
  		Loop radius ($R$)& 0.5 Mm \\
  		Strand radius ($R_\mathrm{s}$)& 0.1$R$\\
  		Magnetic Field ($B_0$)& 7.5 G \\
  		Background density ($\rho_\mathrm{e}$) & $0.5\cdot 10^{-12}$ kg/m$^3$ \\
  		Strand peak density ($\rho_\mathrm{i}$) & $1.25\cdot 10^{-12}$ kg/m$^3$ \\
  		Dense strand peak density & $2.0\cdot 10^{-12}$ kg/m$^3$ \\
  		Background temperature & 2.7 MK \\
  		Plasma $\beta$ & 0.1 \\
  		\hline
  	\end{tabular}
  	\label{table}
  \end{table}
  
\subsection{Boundary conditions}
\label{driver}
In order to model ubiquitous transverse waves, at $z = 0$ we define a velocity driver:
\begin{equation}
\label{driver}
v_y = A\ \mathrm{sin}\left(\frac{2\pi}{T} t\right)\ \mathrm{exp}\left(-\frac{1}{\alpha}\left(\frac{r}{R}\right)^2\right)
\end{equation}
where $A = 5$ km/s is the peak velocity amplitude, $T = 100\ $s is the oscillation period, $t$ is simulation time, $r = \sqrt{x^2+y^2}$ is the radial distance from the center of the loop, and $\alpha$ = 2.5 defines the width of the Gaussian. The velocity amplitude of the driver is somewhat larger than the average observed values for ubiquitous transverse waves of the chosen period (see Appendix~\ref{Appendix} for the results of a simulation with half the amplitude).
The resulting total displacement of the loop structure is $\approx$ 160 km. The other variables at the bottom boundary obey a Neumann-type, zero-gradient condition.   At the top boundary, the propagating waves freely leave the domain, by using the same zero-gradient or outflow boundary conditions on all variables, resulting in an open loop structure. At the four lateral boundaries, we apply the same outflow conditions as for the top boundary: any waves can leave the domain.

\subsection{Numerical method and  mesh}
To solve the 3D ideal MHD problem, we use the \texttt{FLASH} code \citep{2000ApJS..131..273F,2009ASPC..406..243L}, opting for the second-order unsplit staggered mesh Godunov method \citep{2009JCoPh.228..952L,2013JCoPh.243..269L} with Roe solver and \texttt{mc} slope limiter. The code implements constrained transport to keep the divergence of the magnetic field down to round-off errors. An adaptively refined mesh is used, with 3 levels of refinement, for $60\times 60\times 32$ initial numerical cells on a numerical domain of $2 \times 2 \times 50\ \mathrm{Mm}$. This means that the $x-y$ plane is much more resolved ($\sim47:1$) than the $z$-direction, along which we expect the solution to be smooth. The highest refinement cells are thus only 8.3 km in size in the $x-y$ plane, placing strong constraints on the timestep, through the CFL condition \citep{1928MatAn.100...32C}. Simulations with one more level of refinement show more small-scale phenomena, and result in a different deformation of the loop cross-section, but does not alter our main conclusions. In addition, to further strengthen our conclusions, we ran zoomed-in simulations of our model (higher resolution but smaller domain, a so-called ``local box''), which will be described and presented in the `Results and discussion' section. 
 \begin{figure*}
    \centering
    	\includegraphics[width=1.0\textwidth]{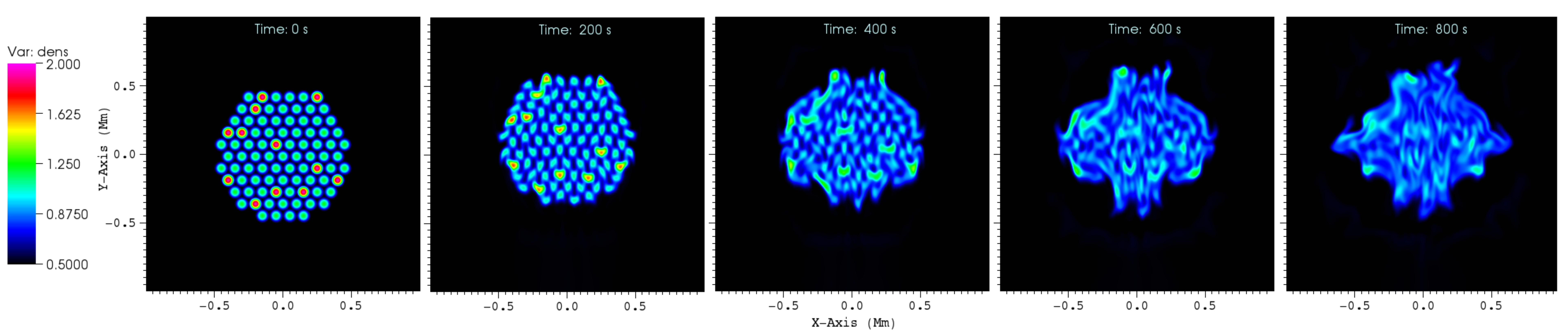}  
    	\caption{Plots of the loop cross-section density, at different times as indicated at the top of each panel. The slices are made at $z = 40$ Mm. The colorbar is common for the plots, in units of $10^{-12}$ kg m$^{-3}$.}
    	\label{img2}
 \end{figure*}
\subsection{Limitations of the model}
The numerical resistivity and viscosity in our code is expected to be orders of magnitude higher than in the solar corona. However, it is not enough to suppress the development of instabilities in our simulations. This is reassuring, since we expect that if instabilities develop in the solar corona, they ensure that large gradients and the energy distribution cascade to smaller scales, where dissipation can be efficient.\par
In our model we neglect effects that do not directly influence the cross-sectional evolution of the loop structure: gravity (density stratification) and geometrical effects (loop curvature). We neglect energy sources and losses (thermal conduction, radiative losses, heating), assuming that our structure is in thermal equilibrium for the duration of the simulation. Again, these effects would have a first order impact on the longitudinal evolution of the loop structure, but here we will focus on the perpendicular dynamics. A realistic solar atmosphere (such as VAL-C) is also missing, as we focus only on the coronal part of the dynamics. Thus, our model is not suitable for the study of the longitudinal evolution of the loop close to its footpoints. \par
Another caveat of this study is the lack of a realistic footpoint driver, opting for a monoperiodic and linearly polarized one in our simulations, for simplicity. This introduces a preferred direction for small-scale formation, as we will see later. 
The chosen internal structure of the loop, as tightly packed parallel, circular strands in a hexagonal arrangement, is of course artificial. We chose this structure inspired by \citet{2013A&A...556A.104P}. Obviously, there are no physical reasons behind this special structuring, and reality will certainly be more complicated. Perhaps the most important question arises about the tight packing of the strands: we believe this to be a valid assumption since the magnetic field in the corona is space-filling due to low plasma $\beta$. Also, this is compatible with state-of-the-art models for multi-stranded loops heated by nano-flares on the individual strands: a small fraction have just been heated, but most of the strands are in the quiescent, cooling phase of their evolution.\\ 
Note also that the cell sizes in our simulations are comparable to the mean free path of the electrons in the corona, or even lower \citep[for values and discussion, see][]{2015RSPTA.37350055P}. However, this does not endanger the applicability of the MHD approximation, thanks to the low plasma $\beta$. 
 
\section{Results and discussion}
\subsection{Whole-loop simulations}
\label{loop}

We run the simulation until a final time $t_\mathrm{f}=900\mathrm{s}=9T$. The evolution of the loop cross-section at $z=40\mathrm{Mm}$ is shown in  Figure~\ref{img2}. It can be seen that the strands, and the overall loop structure, are quickly deformed by the propagating transverse waves. The deformation of strands was previously noted by \citet{2009ApJ...694..502O} using a setup with 4 strands, however, the individual strands could still be clearly identified during their whole simulation. In contrast, in our simulations the initially regular honeycomb structure is completely destroyed and the strands are intermixed to such a degree that we can no longer treat them individually. \par
 The cross-section of the loop structure is a continuously changing and mixing plasma. Smaller structures, down to the numerical resolution, are seen to appear and disappear, as well as longer, highly irregular structures. These deformations and the intermixing appear for two reasons: on the one hand, denser strands, having smaller phase speeds, quickly get out of phase with the neighboring, less dense strands, resulting in ``collisions'' due to phase mixing. On the other hand, individual strands are subject to the Kelvin-Helmholtz instability (KHI) at their boundaries, as previously simulated by \citet{2008ApJ...687L.115T,2014ApJ...787L..22A}. The instability criteria of the KHI for the case of an uniformly rotating plasma cylinder (with discontinuous boundary) can be written as \citep{2015ApJ...813..123Z}:
\begin{equation}
\frac{\rho_\mathrm{i}+|m|\rho_\mathrm{e}}{\rho_\mathrm{i}+\rho_\mathrm{e}}(|m|-1)
\frac{v_0^2}{v_\mathrm{Ai}^2} > R_\mathrm{s}^2 k_z^2 
\label{eq3}
\end{equation}
where $v_0$ is the velocity shear at the boundary, $k_z$ is the longitudinal wavenumber. Equation (\ref{eq3}) suggests that thinner strands are prone to a faster growth of the KHI. Substituting our model parameters, we find that all modes with $m \geq 2$ are unstable. However, this should be taken only qualitatively, given the large difference between the models, in the sense that KHI is expected to occur for our strand parameters.
A key parameter for the development of the KHI and ultimately of the mixing process is the velocity amplitude of the transverse oscillation of the structure. Recently, the displacement amplitudes of decayless kink oscillations were analyzed, and our measured total displacement of the loop structure (0.16 Mm) is well within the range of observed displacements, and very close to the average of 0.17 Mm \citep{2015A&A...583A.136A}. However, the observed periods are on average longer, putting our chosen velocity amplitude at the high end of observed velocities.   \par
\begin{figure*}
    \centering
       \begin{tabular}{@{}cc@{}}
        \includegraphics[width=0.5\textwidth]{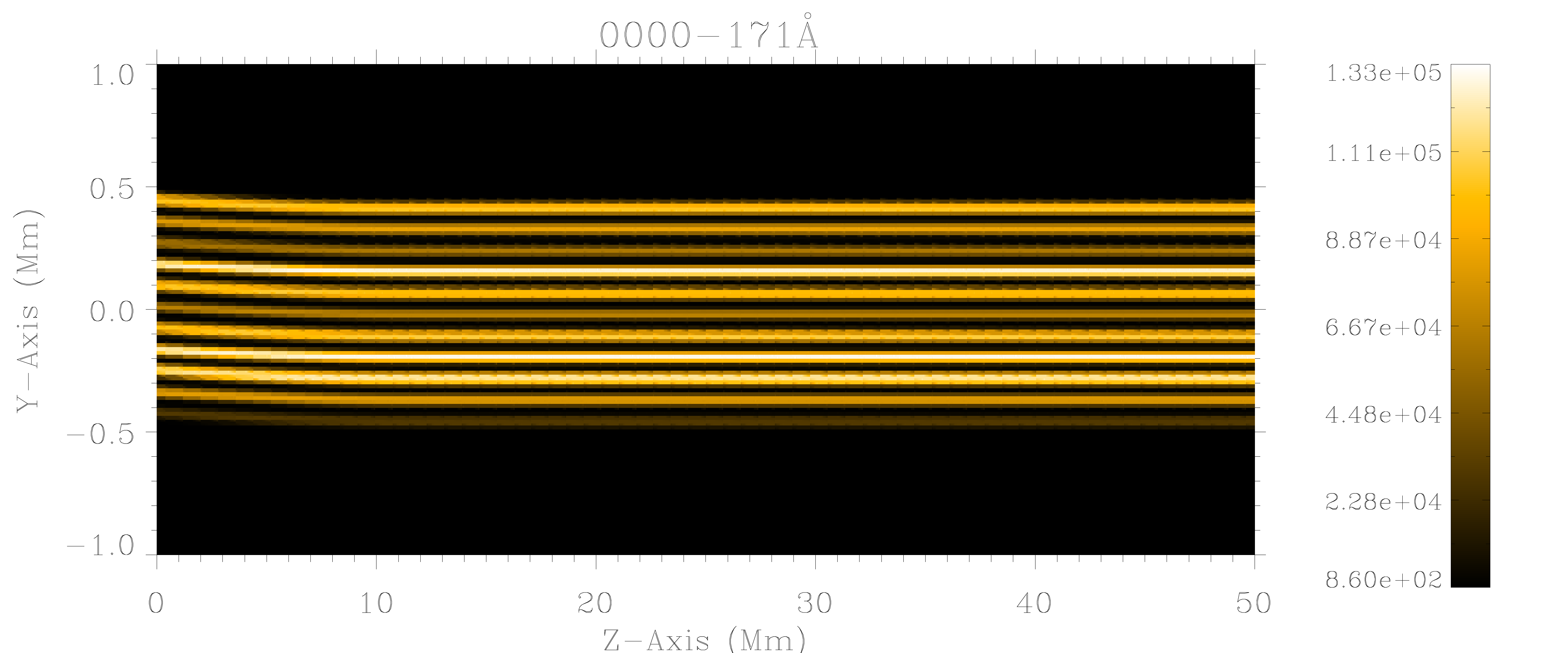}  
        \includegraphics[width=0.5\textwidth]{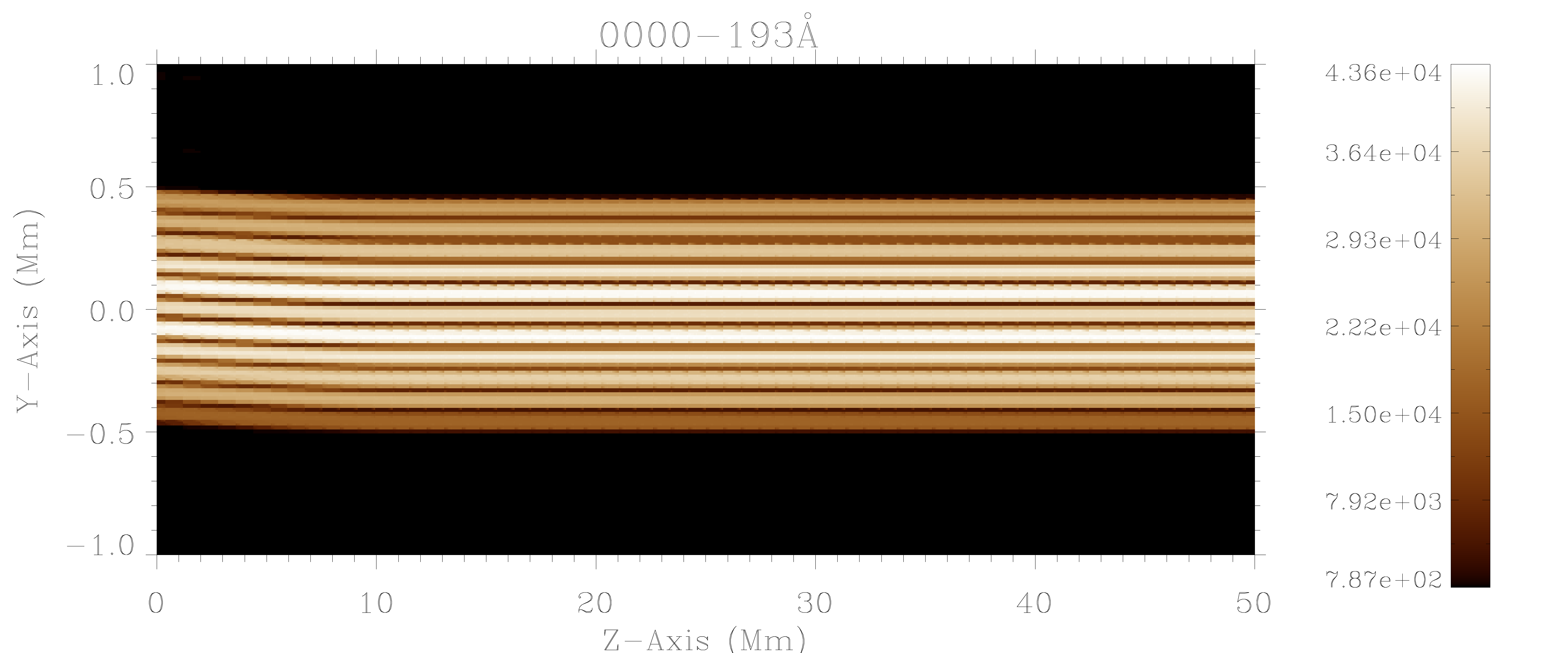} \\
        \includegraphics[width=0.5\textwidth]{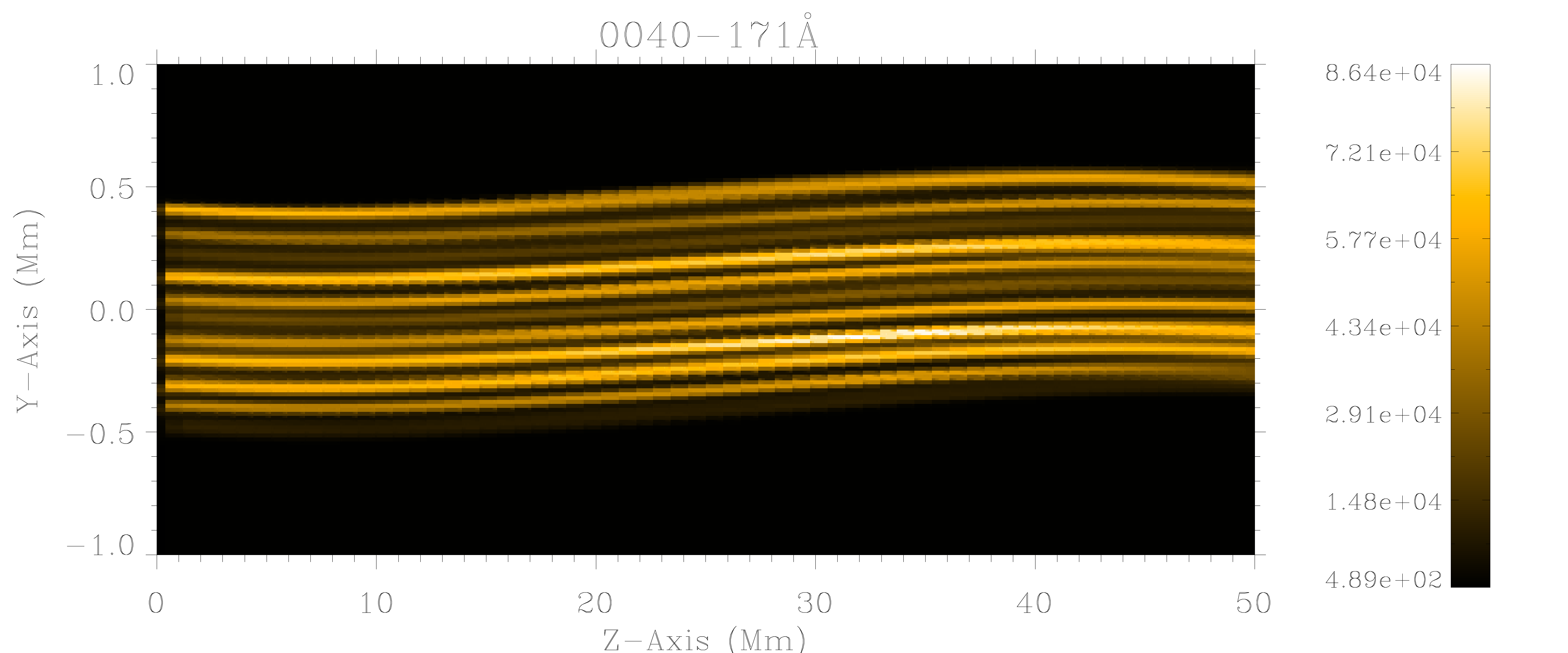} 
        \includegraphics[width=0.5\textwidth]{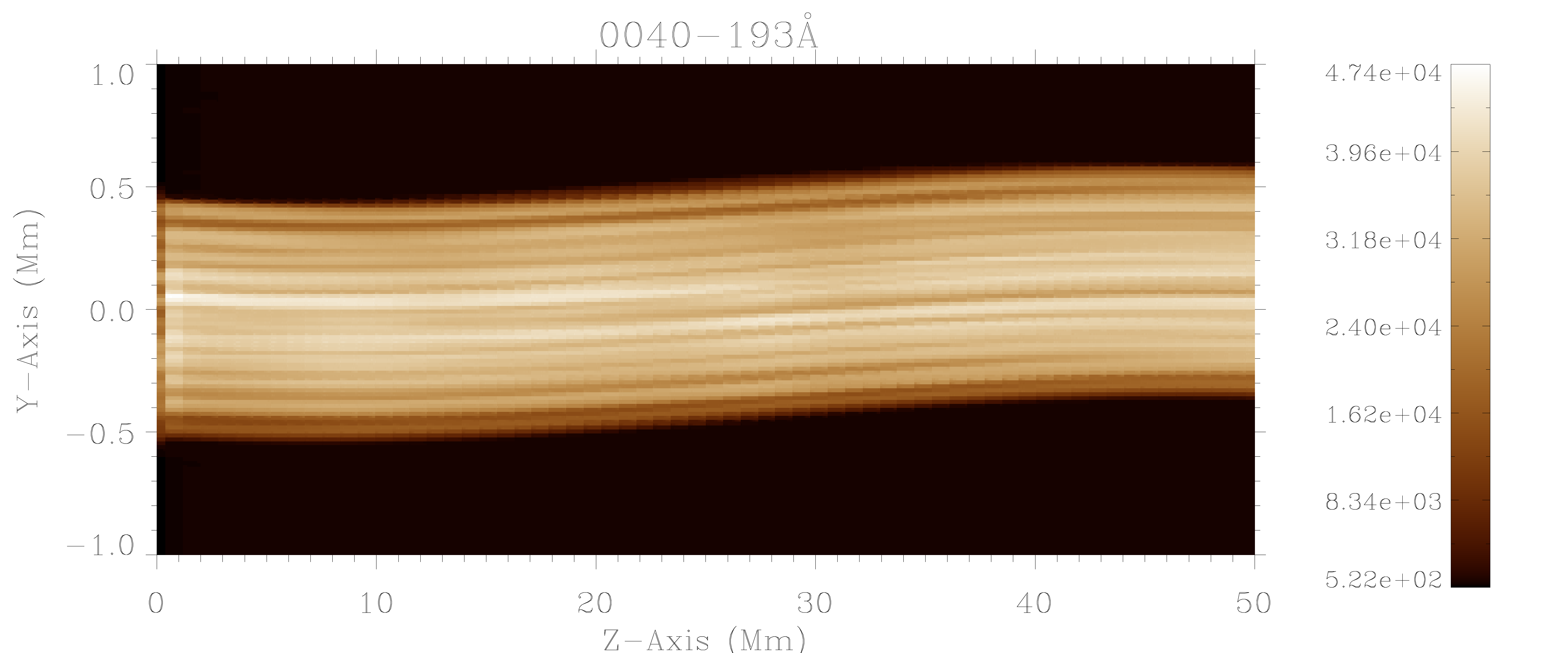} \\
        \includegraphics[width=0.5\textwidth]{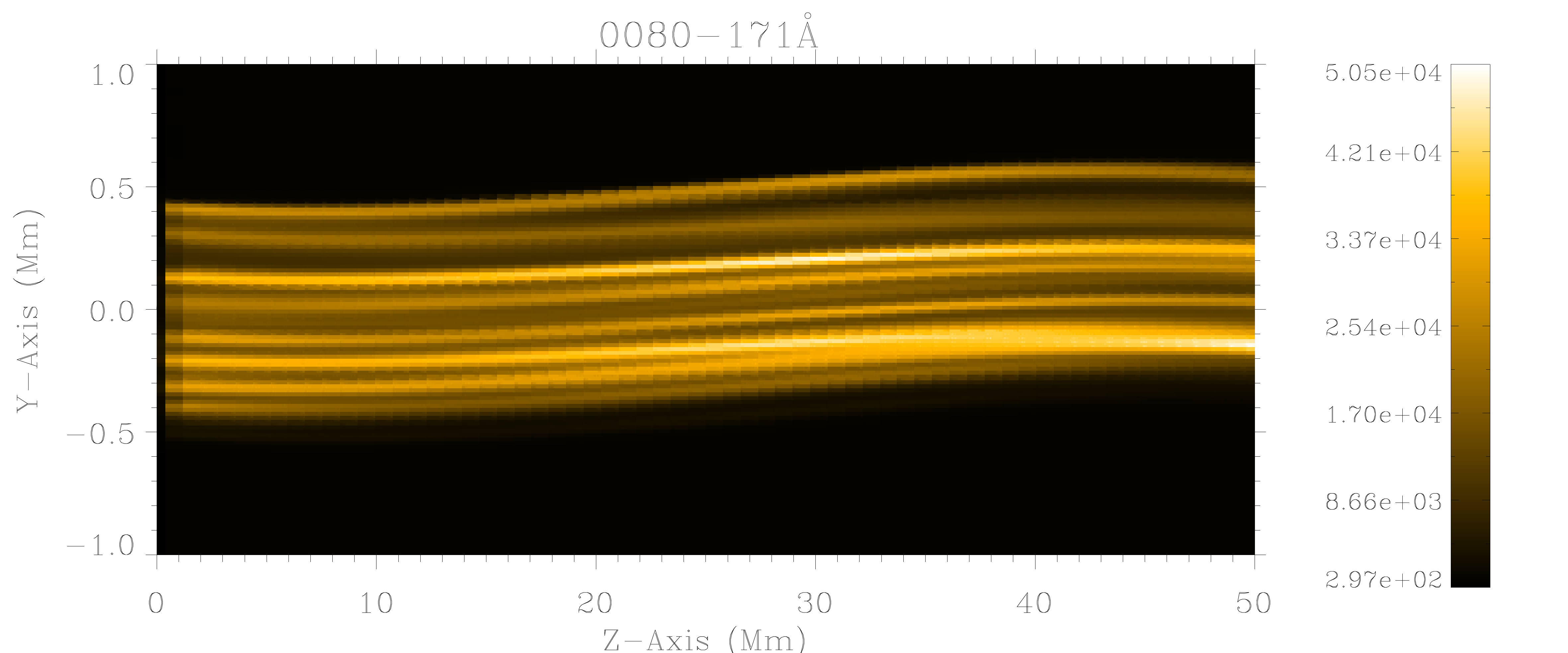} 
        \includegraphics[width=0.5\textwidth]{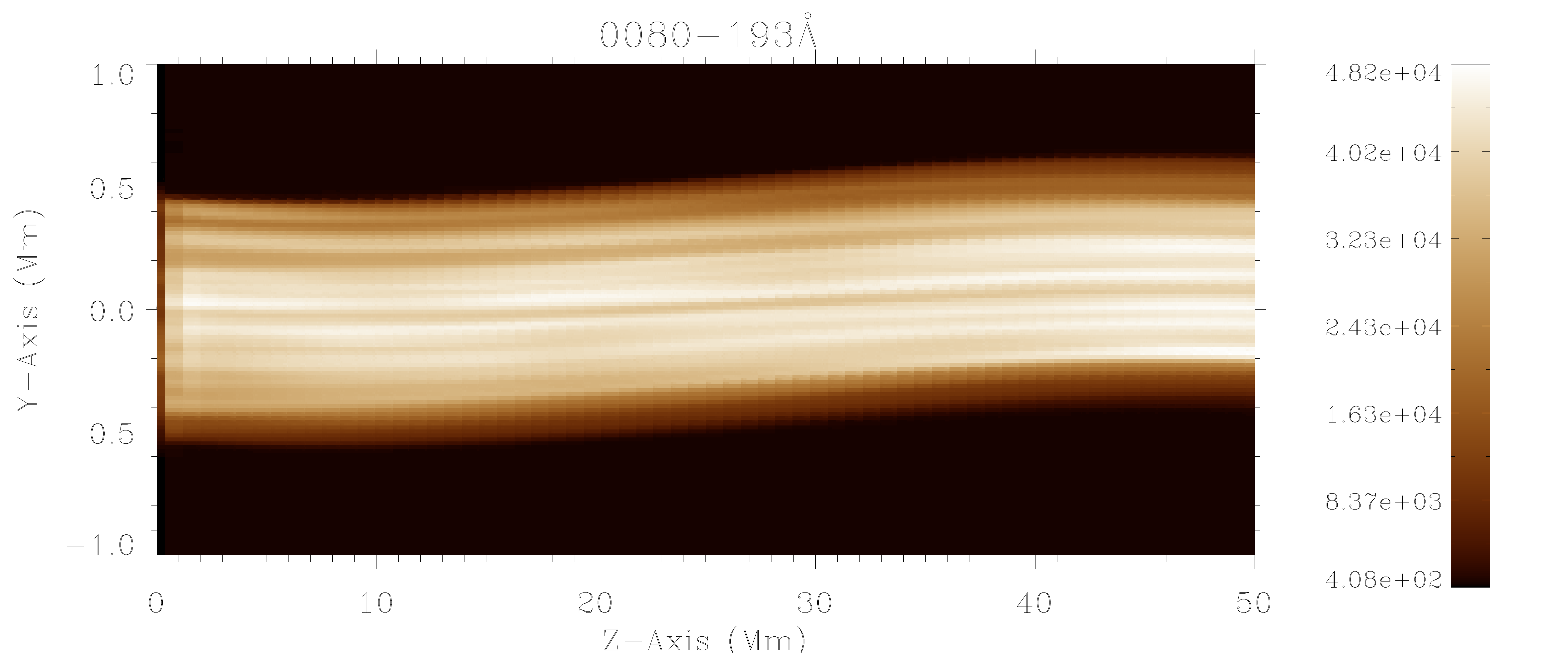} \\
        \includegraphics[width=0.5\textwidth]{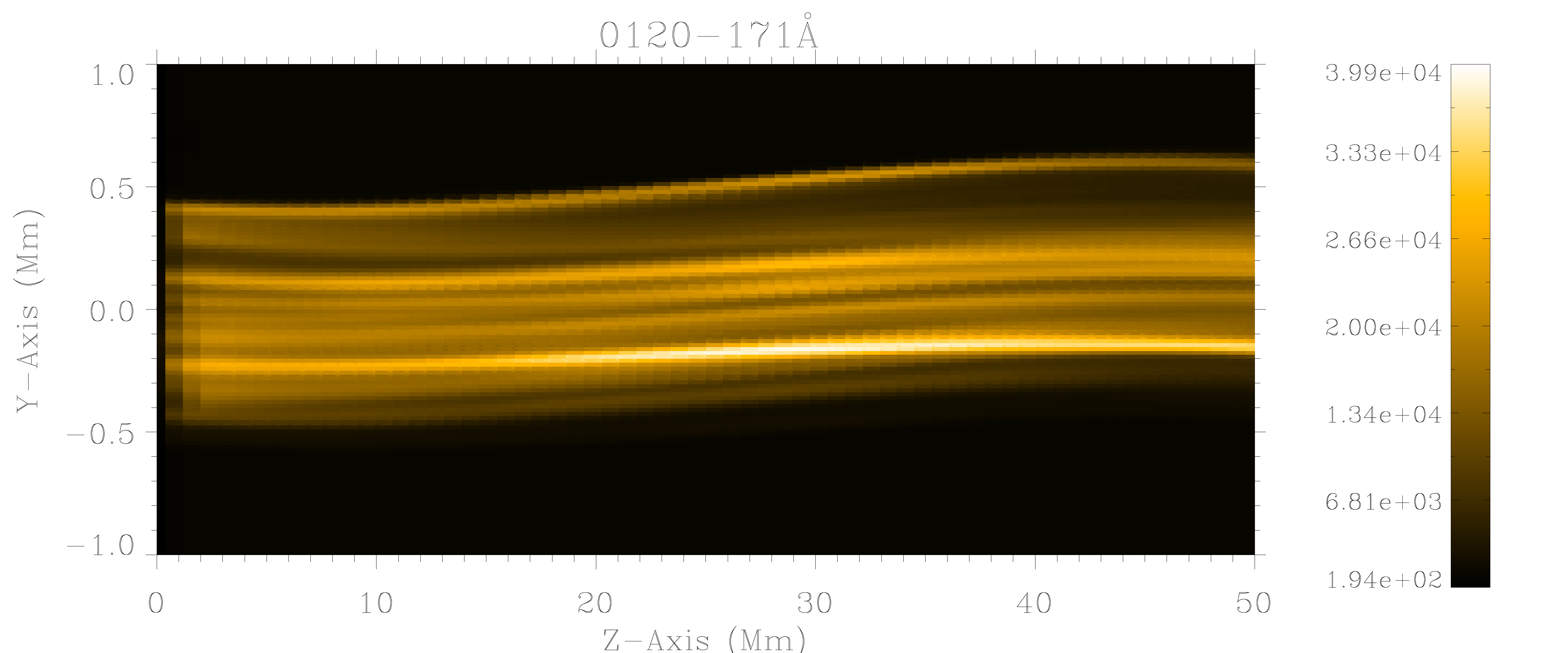} 
        \includegraphics[width=0.5\textwidth]{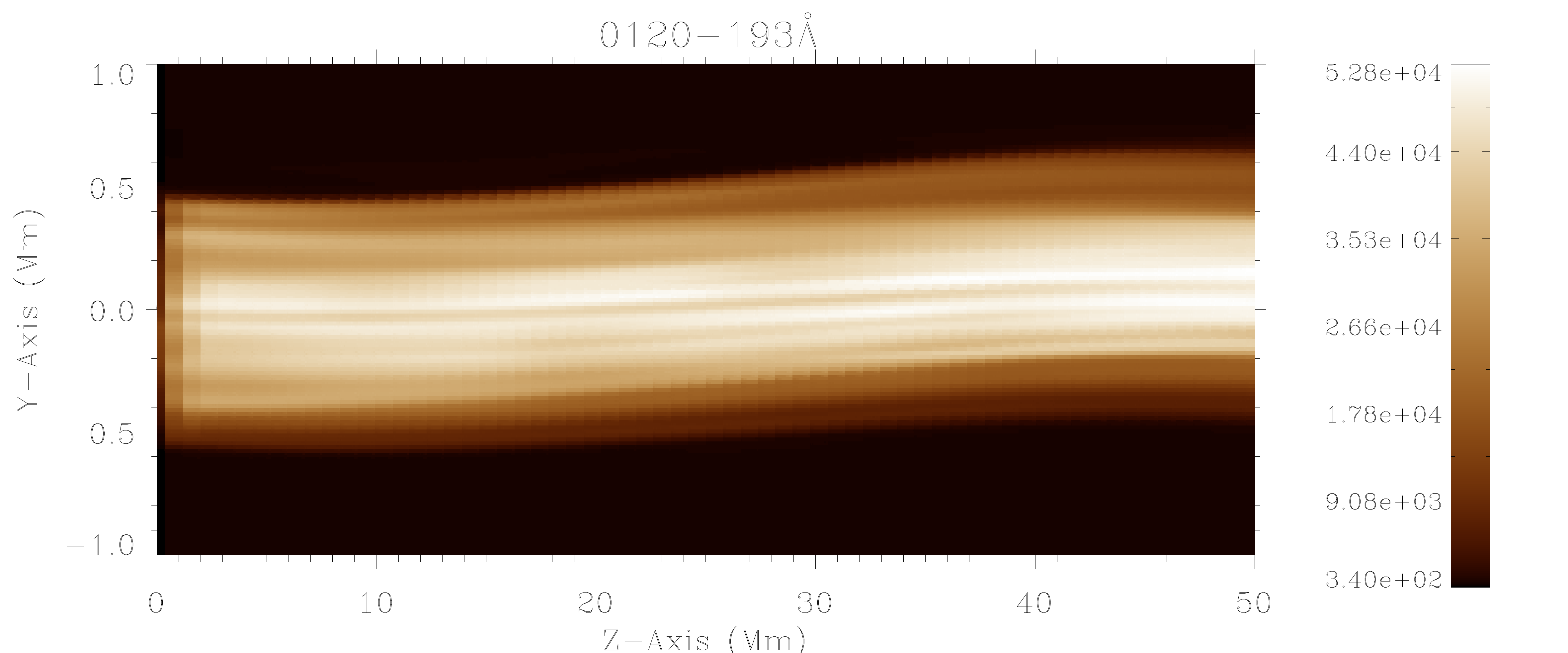} \\
        \includegraphics[width=0.5\textwidth]{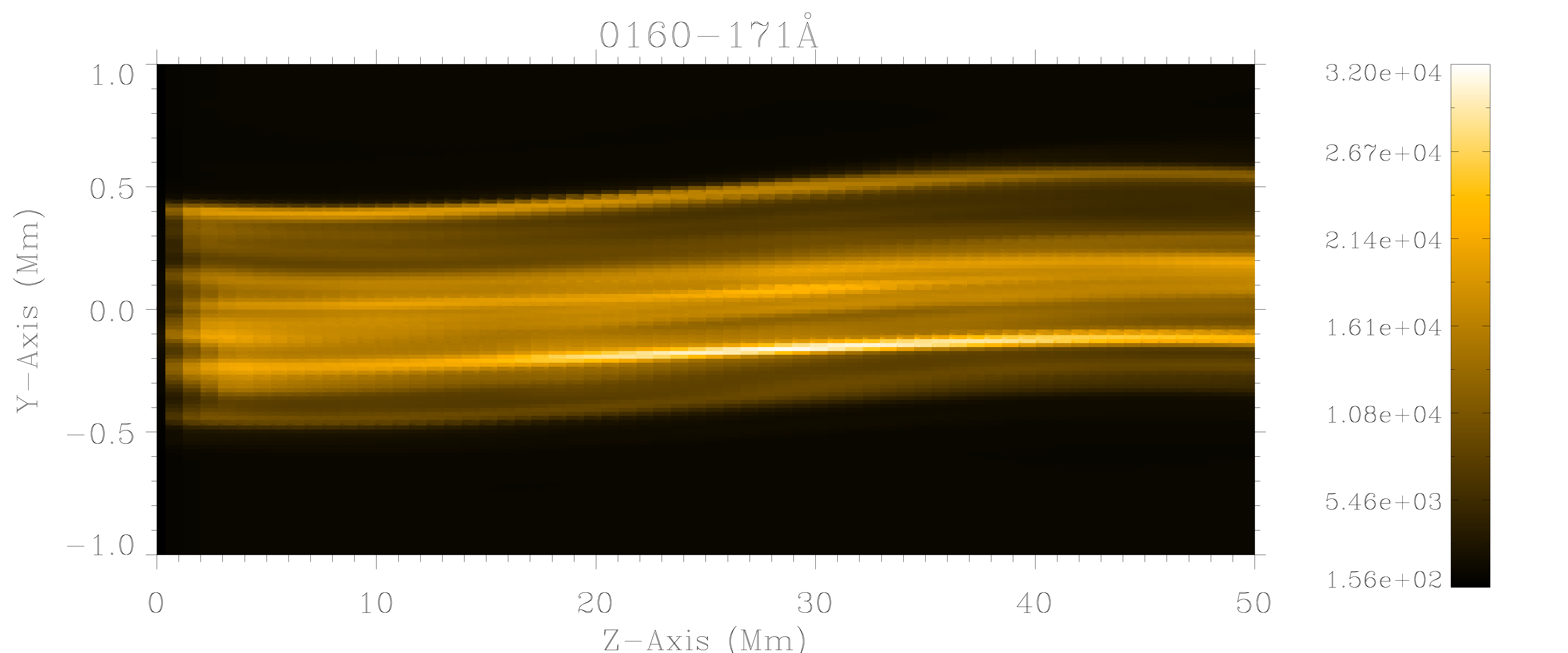} 
        \includegraphics[width=0.5\textwidth]{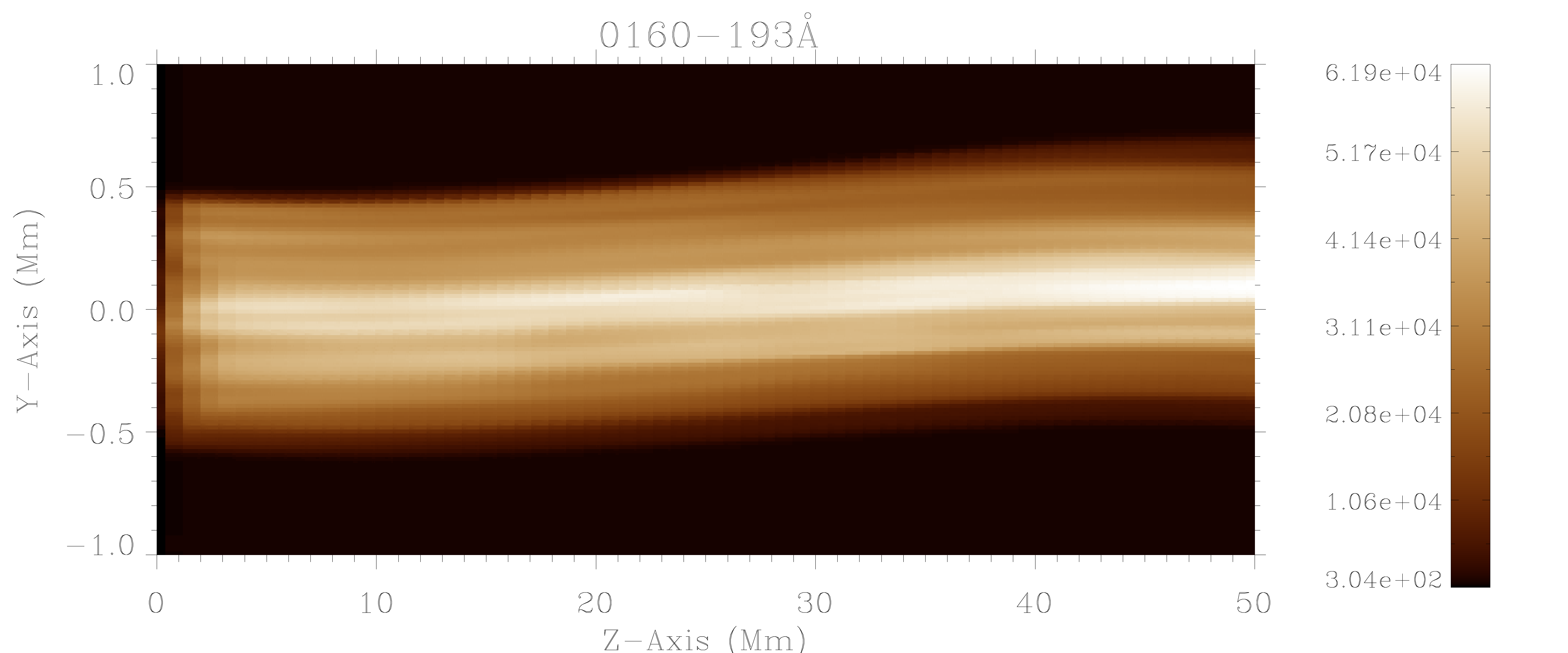} \\
       \end{tabular}  
    	\caption{Synthetic observations of the numerical domain, perpendicular to the $y-z$ plane, with the 171\AA\ (left) and 193\AA\ (right) filters, at different times, corresponding to those in Figure~\ref{img2} (the titles for each plot contain the snapshot number, with one snapshot every 5 seconds). The color bars have units erg cm$^{-2}$ s$^{-1}$ sr$^{-1}$.}
    	\label{img3}
 \end{figure*}
\begin{figure*}
    \centering
       \begin{tabular}{@{}cccc@{}}
        \includegraphics[width=0.25\textwidth]{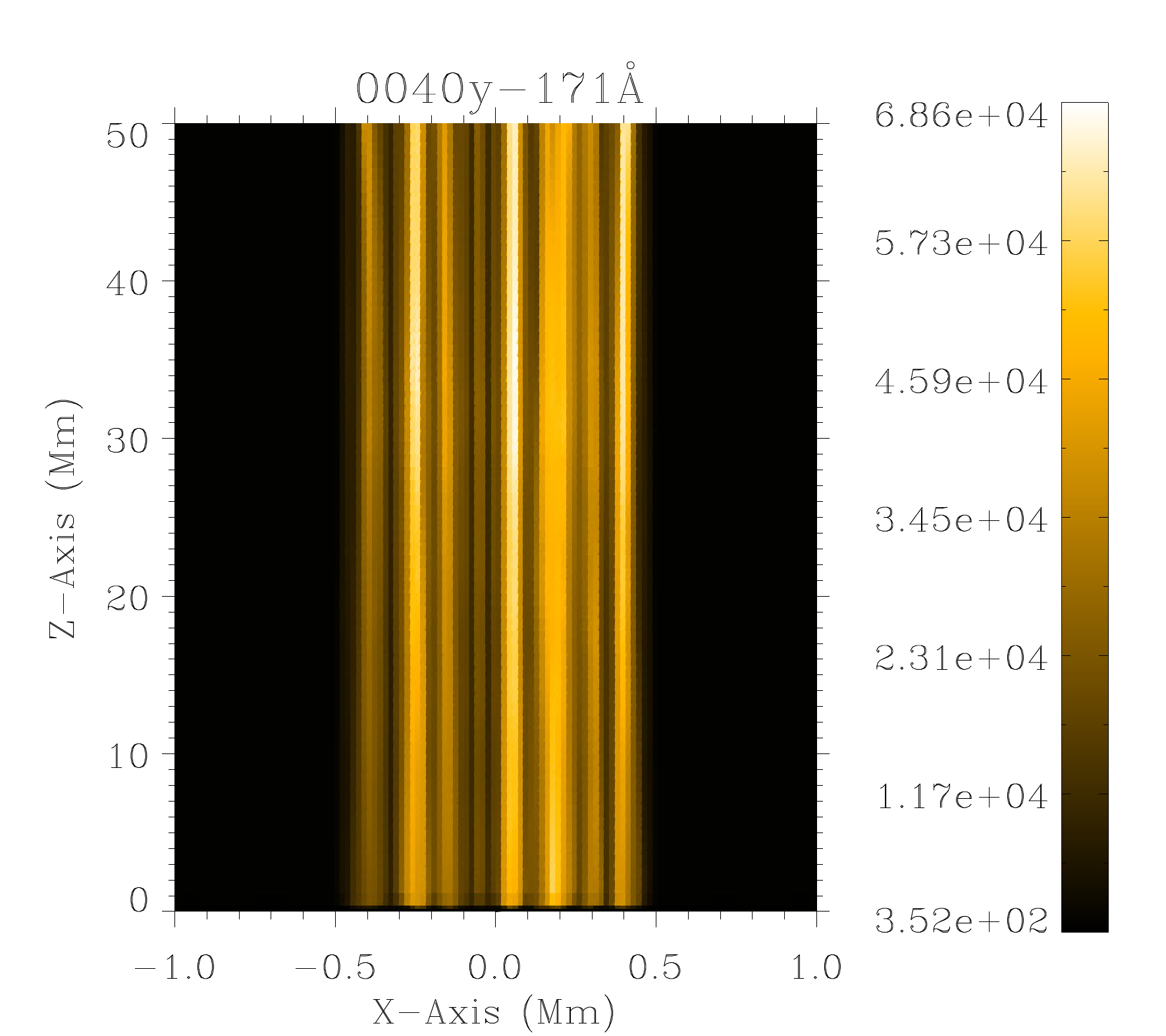}  
        \includegraphics[width=0.25\textwidth]{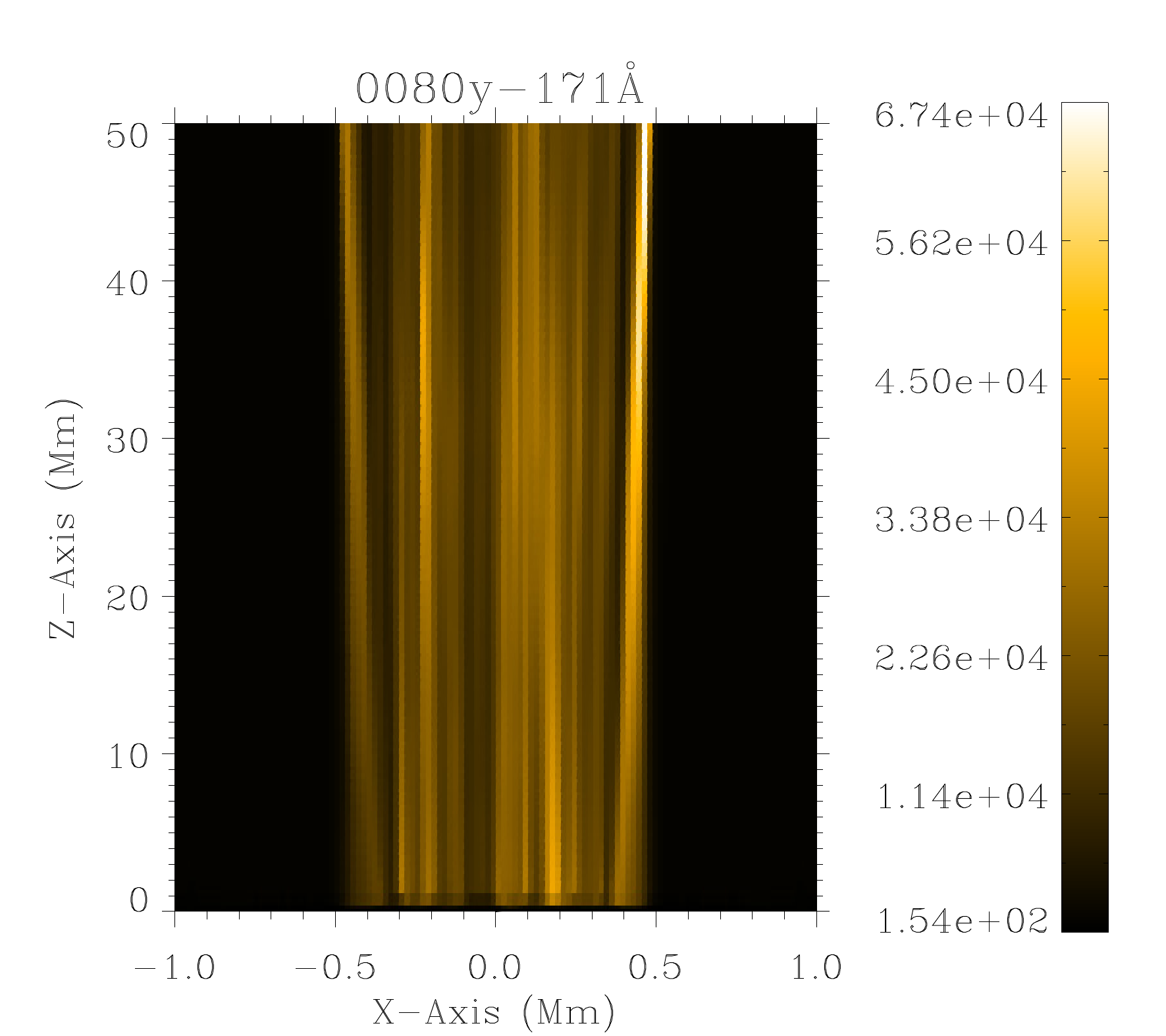}
        \includegraphics[width=0.25\textwidth]{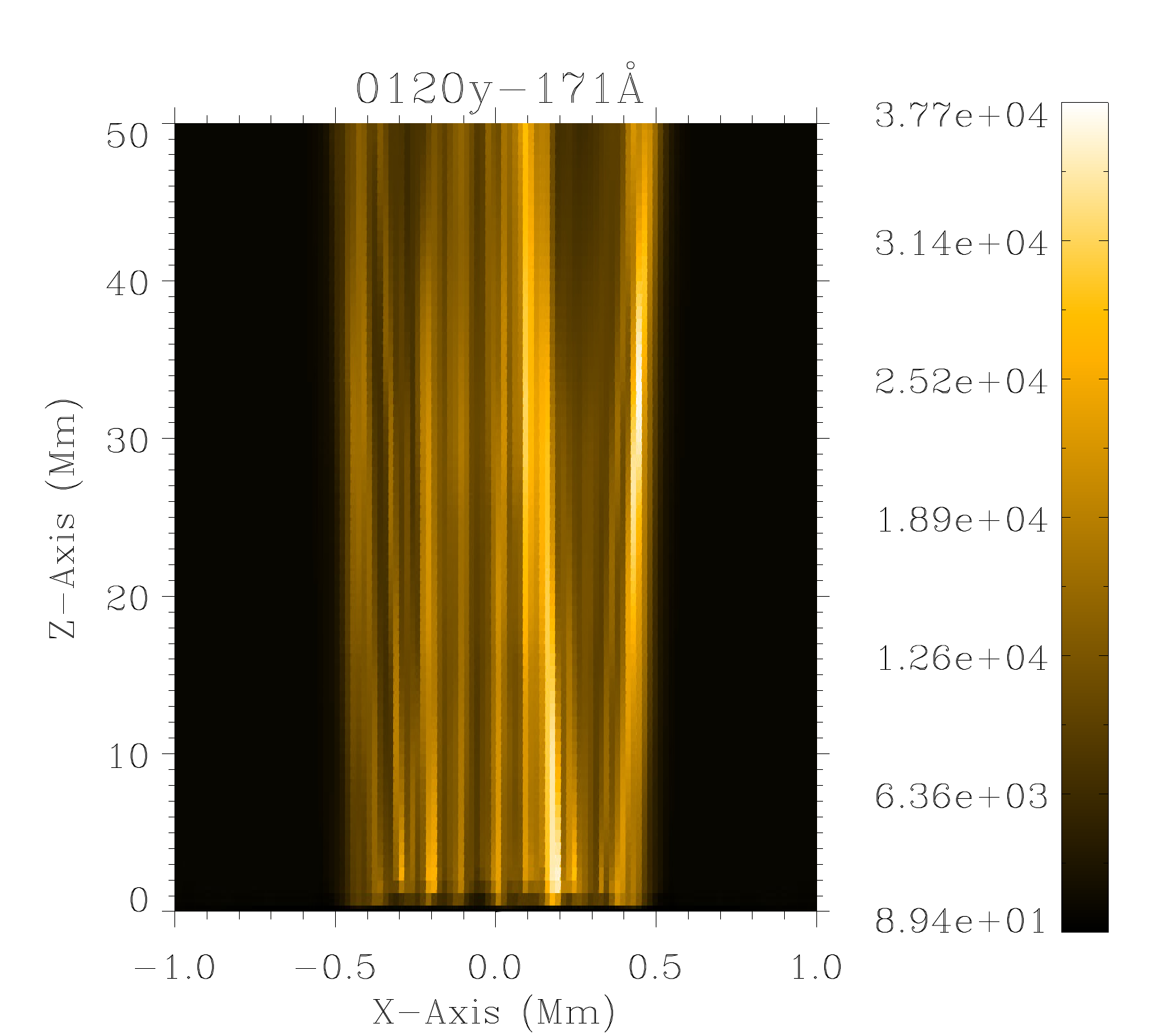} 
        \includegraphics[width=0.25\textwidth]{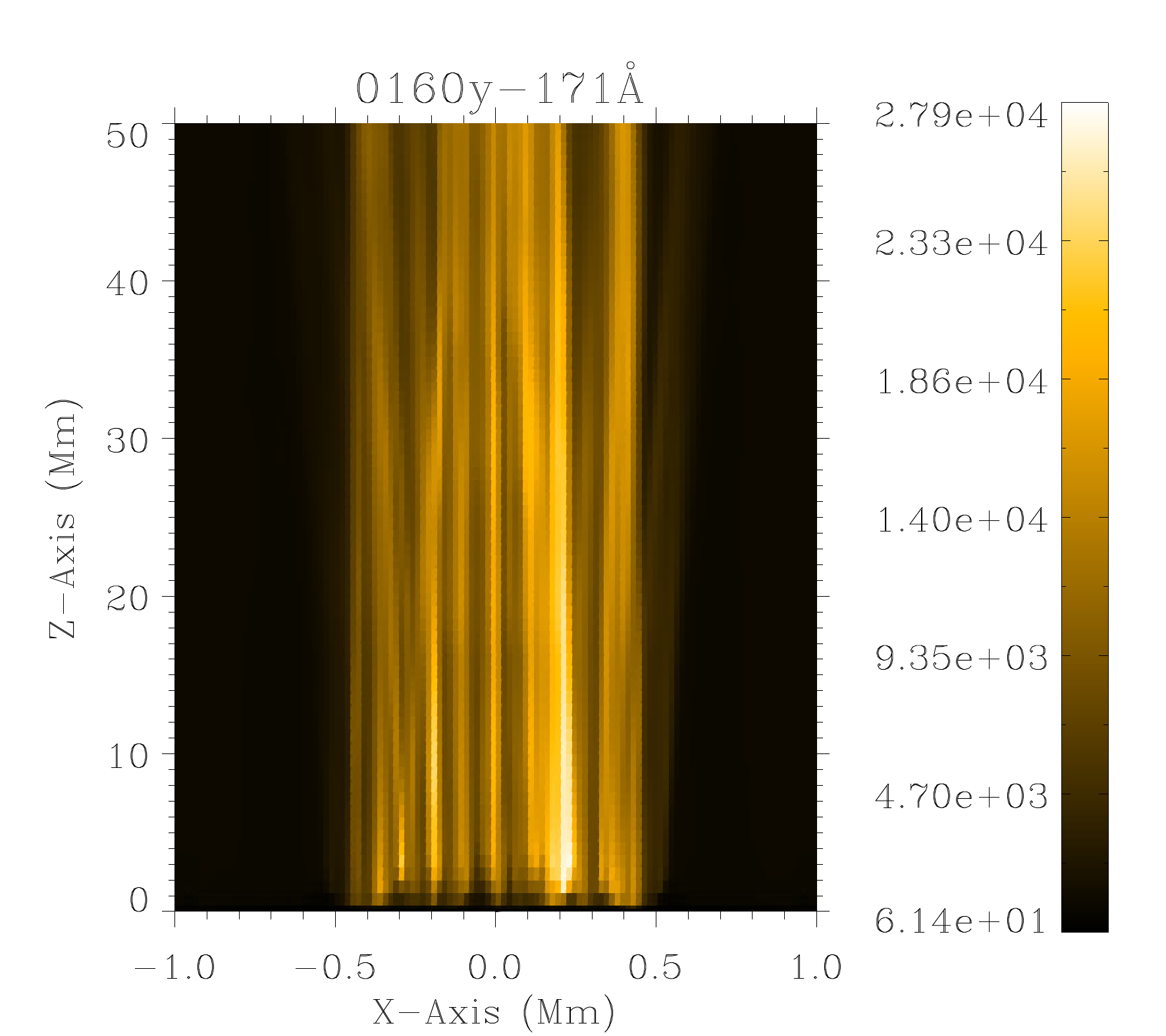}  \\
        \includegraphics[width=0.25\textwidth]{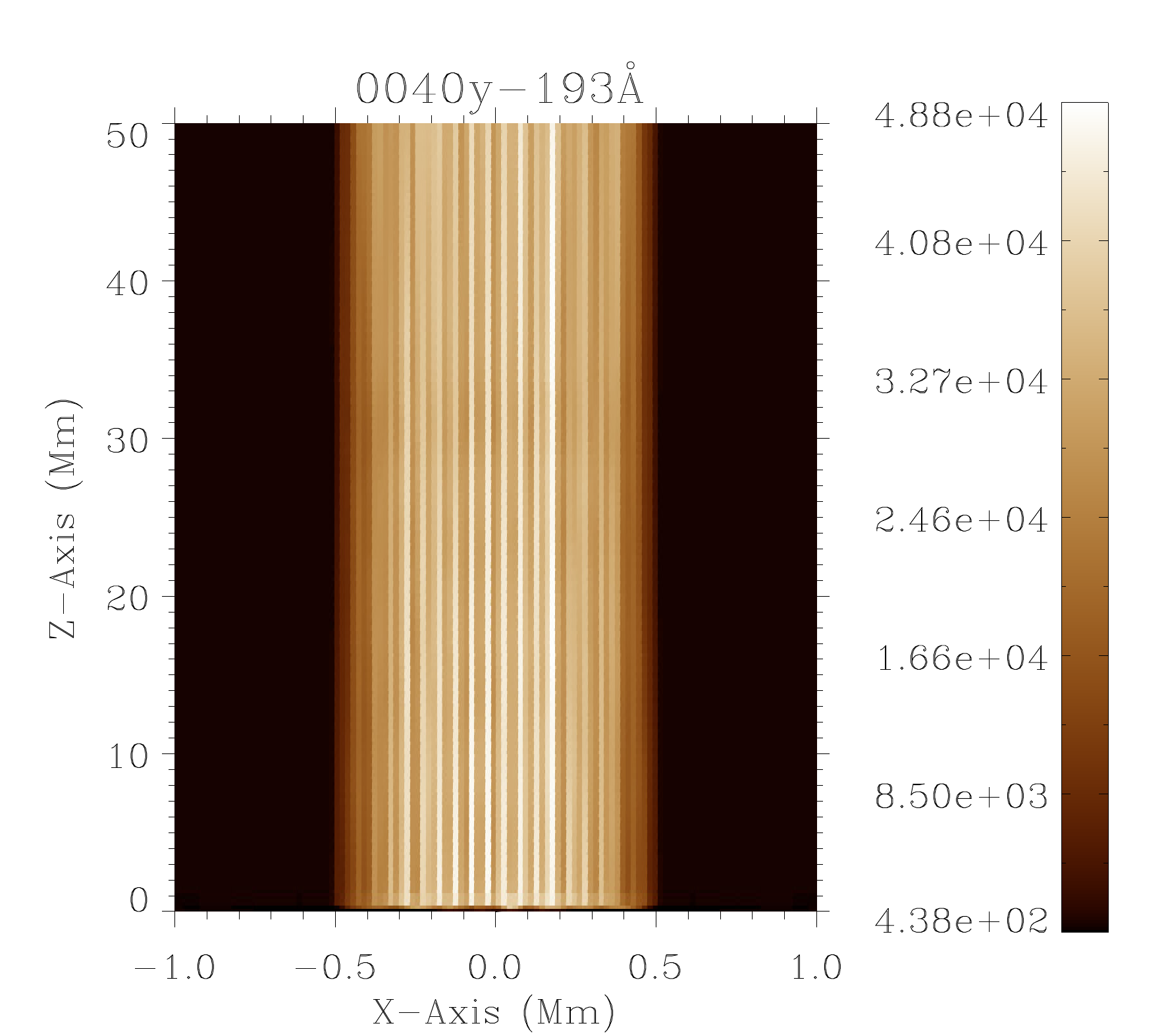} 
        \includegraphics[width=0.25\textwidth]{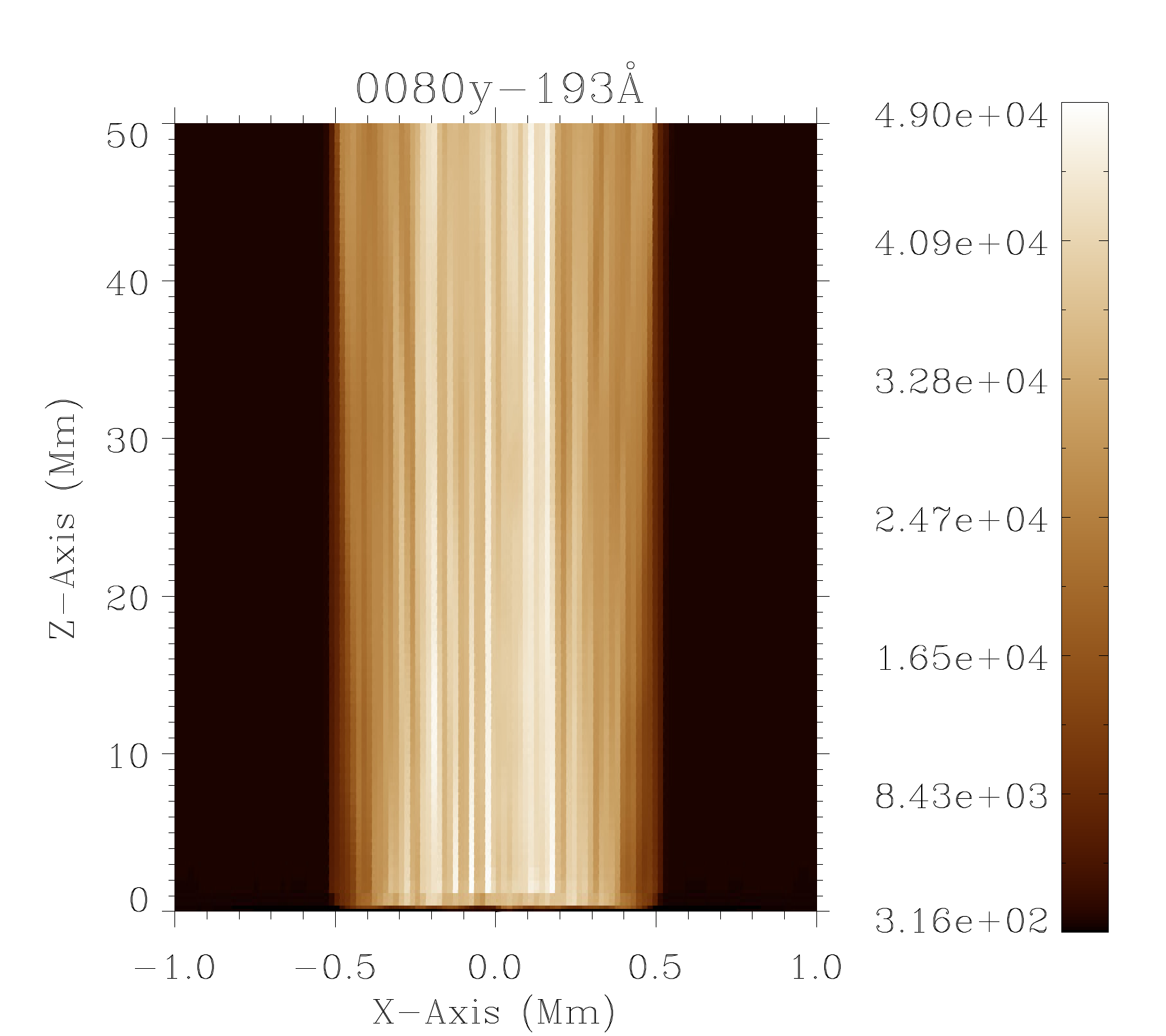} 
        \includegraphics[width=0.25\textwidth]{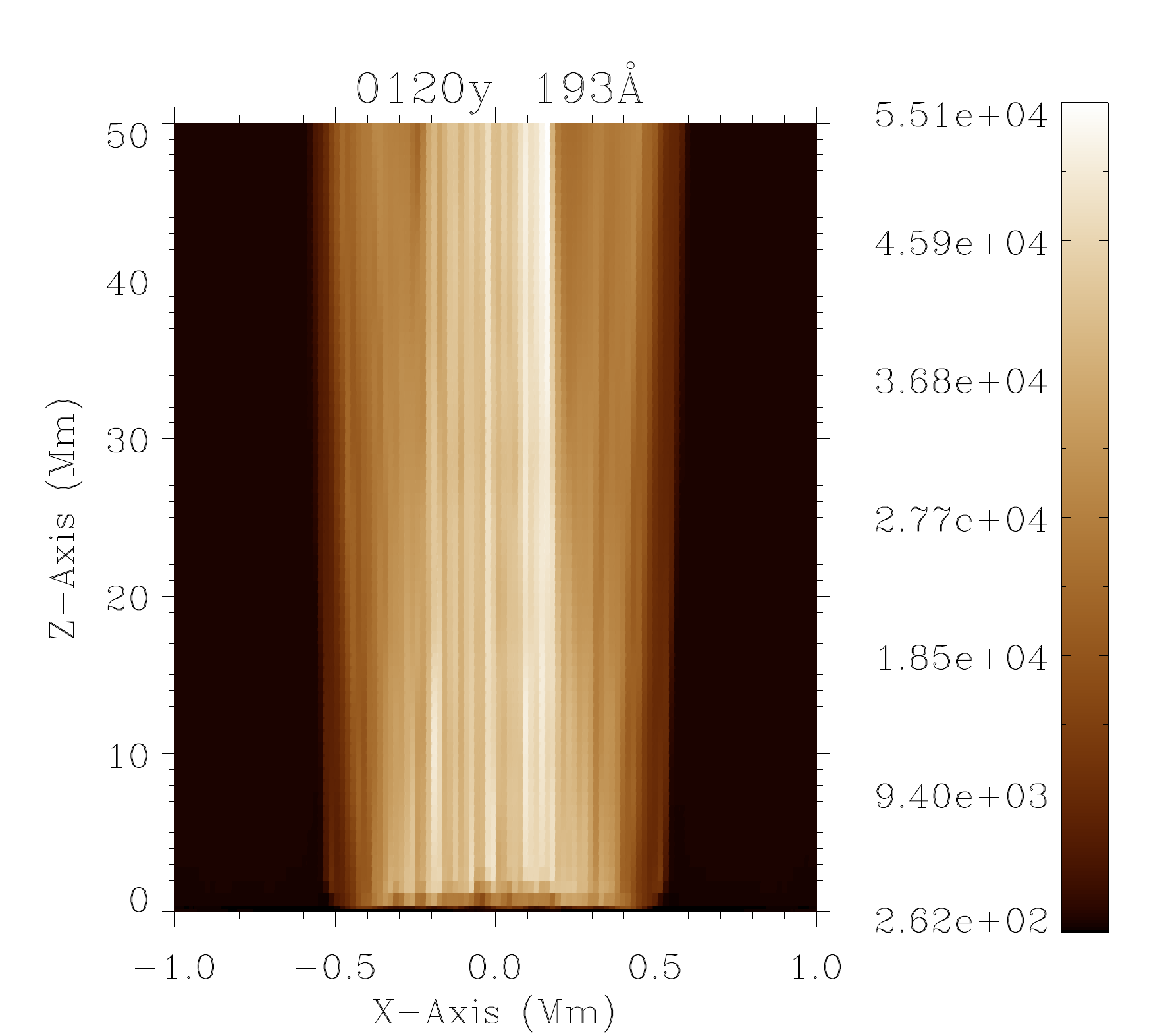}
        \includegraphics[width=0.25\textwidth]{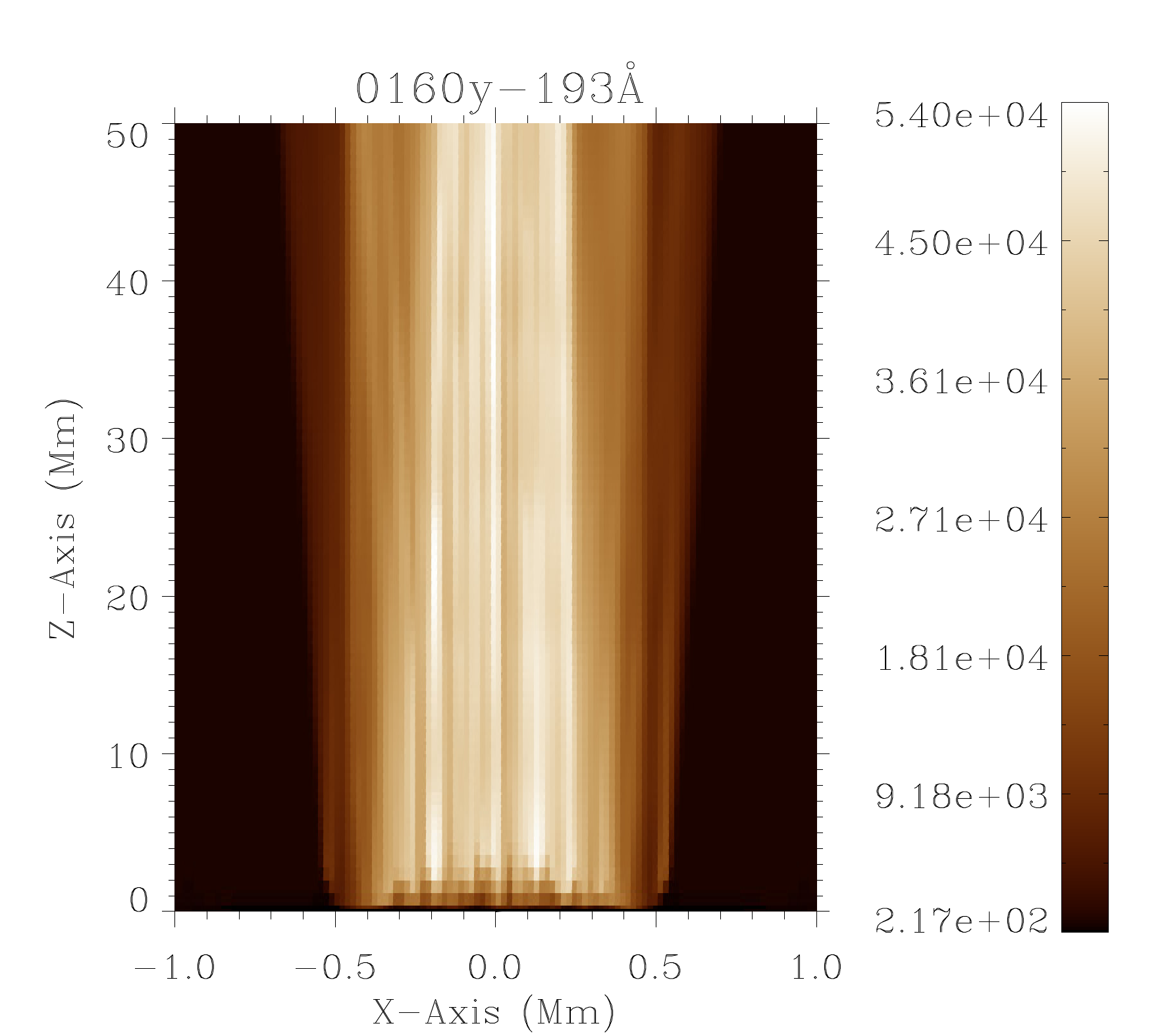} 
       \end{tabular}  
    	\caption{Synthetic observations of the numerical domain, perpendicular to the $x-z$ plane, with the 171\AA\ (up) and 193\AA\ (down) filters, at different times (the titles for each plot contain the snapshot number, with one snapshot every 5 seconds). Note that the $t = 0$ snapshot is not included, and the other snapshots correspond to those in Figure~\ref{img3}. The color bars have units erg cm$^{-2}$ s$^{-1}$ sr$^{-1}$.} 
    	\label{img4}
 \end{figure*}

Synthetic images, corresponding to AIA's 171\AA \ and 193\AA \ filters were calculated for the dataset using FoMo\footnote{https://wiki.esat.kuleuven.be/FoMo/FrontPage} \citep{FoMo}, for two lines-of-sight: along the $x$-axis (Figure~\ref{img3}) and $y$-axis (Figure~\ref{img4}).  
The synthetic images reveal the longitudinal evolution of the loop structure. We can see that the conclusions on Figure~\ref{img2} hold also here: the regular, clearly-stranded structure is destroyed by the mixing: strands disappear and new strands appear. The newly formed strands have a well-defined longitudinal appearance, and can generally be clearly identified along the whole length of the loop. This matches well with the observations. However, the newly formed, apparent strands in the later stages of the simulation are not strands in the literal sense of the word, because it is clearly seen in the right panel of Fig.~\ref{img2} that all plasma in the loop is mixed.\par 
The emission in the 171\AA\ line is dominated by the denser strands of our loop, because they have a lower average temperature of 0.8 MK. The maximum intensity in this channel is decreasing in time (by a factor of four towards the end of the simulation $t_\mathrm{f}$), indicating that the dense strands diffuse and mix with hotter plasma. This also causes the intensity in the 193\AA\ channel to steadily rise, albeit slower. To be noted here also is the fuzzier appearance of the loop in the 193\AA\ channel. In the images generated with a line-of-sight along the $y$-axis (Figure~\ref{img4}) more and finer small-scale structures are seen in both passbands. This can be explained by the specific linear polarization of the boundary driver: the strands tend to get elongated in the direction of the displacement, and the KHI occurs at the boundary of the strands perpendicular to the direction of motion. \par
\subsection{Column emission measure}

\begin{figure}[h!]
    \centering
    	\includegraphics[width=0.5\textwidth]{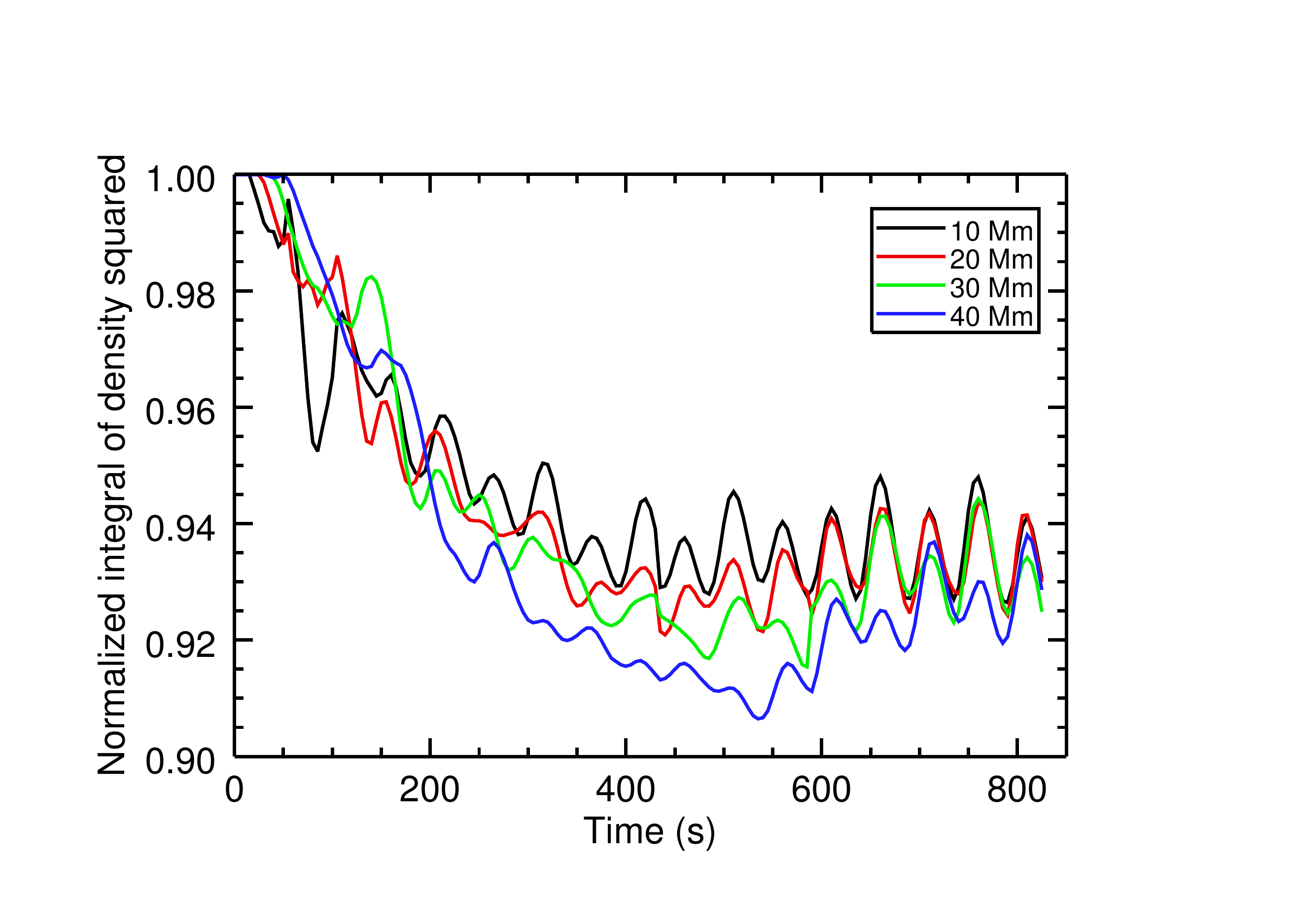}  
    	\caption{The evolution of normalized column emission measure at four different heights ($z$-axis).}
    	\label{filling}
 \end{figure}

We now calculate the column emission measure, which is proportional to the volume filling factor of the plasma \citep{1995ApJ...454..499P,2015ApJ...800..140G}, at different heights of the loop structure (Figure~\ref{filling}). It is defined as
\begin{equation}
EM = \int_{LOS} n_e^2 dy
\end{equation}
where $n_e$ is the electron number density of the optically thin plasma, and the integral is over a line-of-sight (LOS) along the $y$-axis. As we are interested only in the change of this quantity over time, we replace $n_e$ by the density of the plasma.
We consider volumes (3D boxes) through the loop structure with LOS area corresponding to AIA per pixel spatial scale (0.6$''$), the surface normal being parallel to the $y$-axis. Thus, we compute, at a specific height $z$ and $x=0$ 
\begin{equation}
EM = \int_{-d/2}^{d/2} dx \int_{z-d/2}^{z+d/2} dz \int_{LOS} \rho^2 dy 
\end{equation}
where $d = 0.43$ Mm.
 
The column emission measure initially decreases at all heights by a small extent (6-8\%), but it ultimately oscillates around a constant value. We can state that the column emission measure, for practical purposes like observational determination of the filling factor, is only slightly changing in the process of loop cross-section mixing. Thus, even though the later stages of the simulation have very mixed plasma, still the observations would yield a similar filling factor as in the classically multi-stranded case.
\subsection{Distribution of structure lengthscales}
\begin{figure}[h!]
    \centering
    	\includegraphics[width=0.5\textwidth]{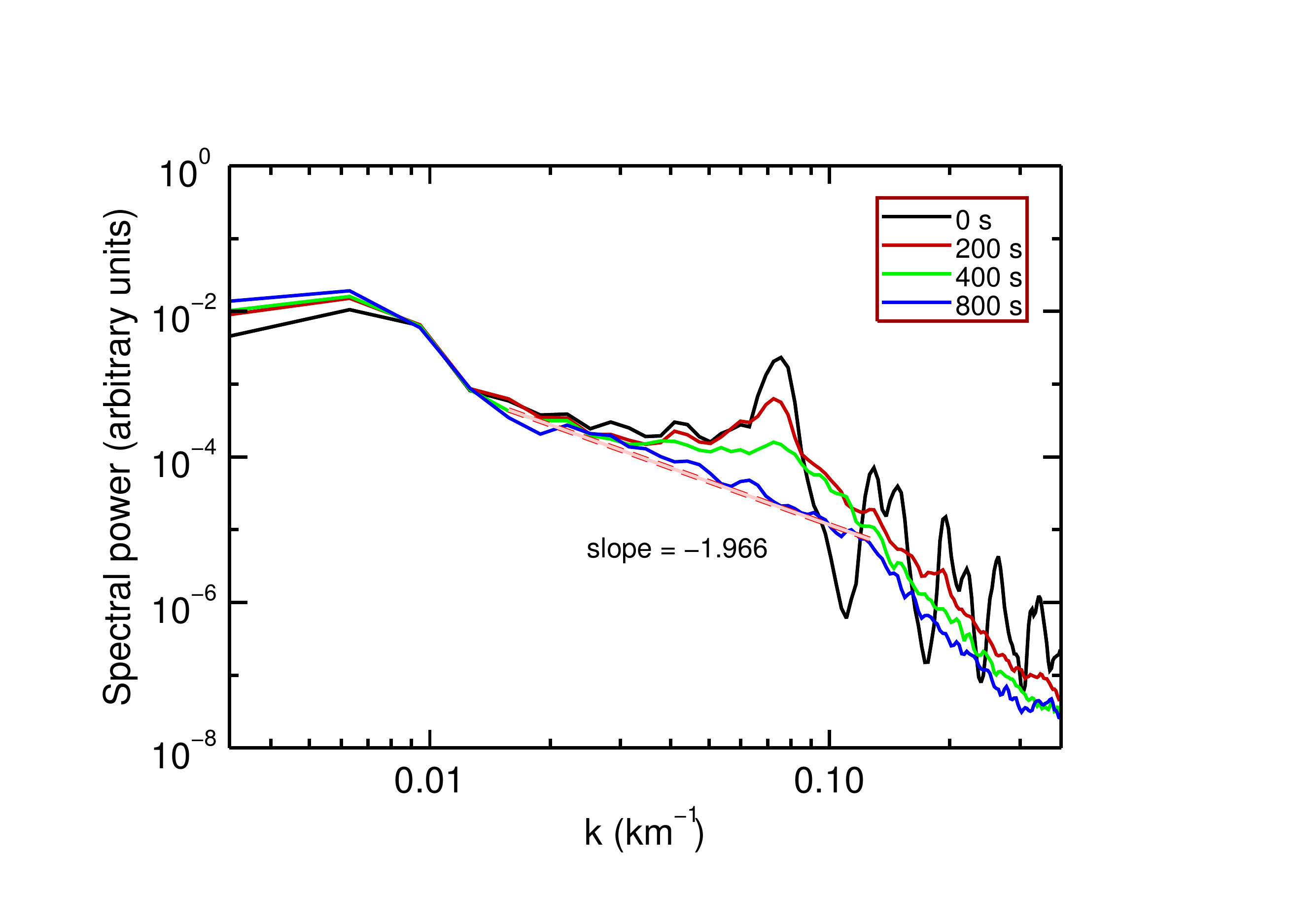}  
    	\caption{Power spectra for the density cross-section of the loop structure at $z=40$ Mm at four different times. Also shown is the linear fit to the central part of the power spectra at $t = 800$ s.}
    	\label{spectra}
 \end{figure}

In order to quantify the change in the distribution of spatial scales over time, we Fourier analyze the density in the cross-section of the loop structure over time. We expect relevant spatial scales to appear as a peak in the power spectra, as it is the case with the specific radius of the strands (see Figure~\ref{spectra}). To obtain this figure, we integrate the k-space of the power spectra as a function of $k = \sqrt{k_x^2+k_y^2}$, where k is the wavenumber (inverse of spatial scale, with a factor of $2\pi$). 
 
This analysis demonstrates that, due to transverse oscillations the internal structure of the loop shows progressively less discrete structuring (i.e. strands of well defined radius), replacing it with a power-law like distribution of scales. The power law index of the inertial range at $t = 800$ s in our simulation is -1.966. In other words, the initially ordered interior of the loop structure transitions into a turbulence-like cross-section.

\begin{figure*}
    \centering
    	\includegraphics[width=1.0\textwidth]{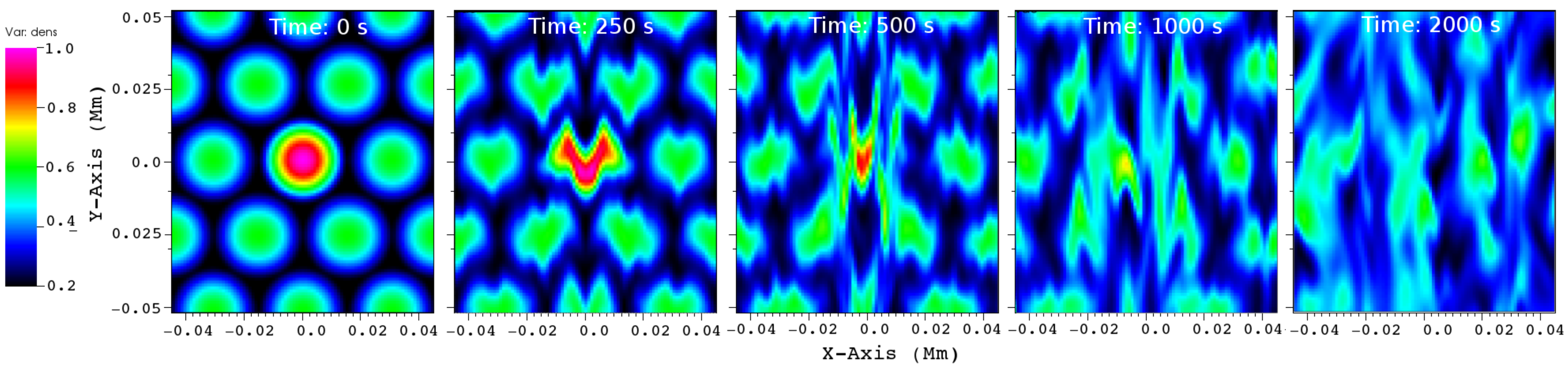}  
    	\caption{Plots showing the density in the $x-y$ plane (cross-section of the strands), at times shown on the top of each plot. The slices are made at $z = 10$ Mm. The color bar is common for the plots, in units of $10^{-12}$ kg m$^{-3}$.}
    	\label{img5}
 \end{figure*}
 
\subsection{Local-box simulations}

In order to strengthen our conclusions, we explore how a bundle of strands with even smaller radius evolve when driven by a continuous transverse driver. We ran simulations with strands covering the entire numerical domain, which can be viewed as a subset of our loop structure in the previous section (using different parameters, however). For these simulations we use the \texttt{MPI-AMRVAC} code \citep{2012JCoPh.231..718K,2014ApJS..214....4P}, with the three-step \texttt{HLLC} solver, in order to solve the 3D ideal MHD problem. To make the computation less demanding, we lower the background and peak densities to $\rho_e = 0.2\cdot 10^{-12}$ kg/m$^3$ and $\rho_i = 0.6\cdot 10^{-12}$ kg/m$^3$, respectively. The dense strand has a peak density of $\rho_i = 1.0\cdot 10^{-12}$ kg/m$^3$. We also adjust the boundary driver in the following way:
\begin{equation}
v_y = \mathrm{A\ sin} \left( \frac{2 \pi}{T} t \right)
\end{equation}
where A$=2.5$ km/s and $T = 100$ s. While previously the driver had a localised shape, the new velocity profile does not depend on spatial coordinates: the whole boundary is forced equally, mimicking the smaller scale of the simulated domain compared to the size of the driver. \\
We switch the lateral boundaries to periodic boundary conditions in order to approximate a larger system. In the boundary opposing the driving boundary we impose the same conditions as before: waves leave the domain freely. We also reduce the size of the strands ($R_\mathrm{s} = 15$ km), to get closer to the observational constraint on their radius \citep{2013A&A...556A.104P}. The physical size of the numerical domain is $0.09 \times 0.104 \times 10$ Mm, being slightly larger in the $y$-direction in order for the strands to overlap perfectly at the (periodic) boundaries of the honeycomb structure. The numerical resolution is uniform throughout the $x-y$ plane, $~1$ km per cell. The resolution in the $z$-direction is 312 km per cell. Simulations with double the resolution show even smaller scale structures and a higher fragmentation of the strands. \par
Comparing Figure~\ref{img5} with the cross-sectional evolution of the previous model (Figure~\ref{img2}), we can spot similarities: The individual strands initially deform, being affected by the KHI at their edges, then fragment, break up or unite, forming new, both smaller and larger scale structures. In this specific example, the evolution of the dense strand can be followed. The dense strand is initially affected by the KHI and the interaction with the other strands, and it becomes highly fragmented in the process. In this violent shaking (having in mind the size of the strands, even the low amplitude ubiquitous transverse waves can be considered as such), the strand eventually breaks up and blends into less dense strands, forming new small scale structures. Ultimately, the other strands tend to intermix, the system displaying a tendency to homogenize the cross-section, losing all identification of loop strands.\par
These local-box simulations represent a sub-section of the simulation in Sec.~\ref{loop}, and here as well the result of the previous simulation is confirmed. Multi-stranded structures cannot keep existing in a perpendicularly driven environment and mix efficiently with their surrounding.

\section{Conclusions} 
       
In the present paper, we aimed to study the dynamics of a multi-stranded coronal loop when subjected to a continuously driven transverse motion. The loop was initially composed of closely packed thin strands in a honeycomb structure. The strands had different densities and temperatures, representing a small filling factor as observed in the corona. The driving velocity amplitudes were close to the amplitude of the observed ubiquitous transverse motions. Such a set-up was studied on a large scale, but also in a local-box simulation.\par
We have shown that the internal multi-strand structure of the coronal loop is quickly destroyed due to instabilities present in individual strands and inter-strand collisions. The strands intermix, forming new small scale structures, both larger and smaller than the size of the original strands. In this case, the approximation that individual, neighbouring strands have an independent hydrodynamic evolution breaks down. This result has direct implications on how coronal loops can be modeled. It poses a strong question on the applicability of multi-strand loop models, given that they are unstable for a transverse driving. \par
The later stages of our simulation show that, if loops are multi-stranded by nature (e.g. by localised heating by nano-flares), their cross-sectional internal structure should be regarded as dynamic and continuously changing, with substructures appearing and disappearing on a large range of length scales. Even in the later stage, the simulations match closely to the observations, with apparent strands and a filling factor which is more or less conserved during the entire simulation run. \par
The results presented here indicate (and simulations not presented here confirm), that in a large amplitude oscillation event of a multi-stranded coronal loop, e.g. induced by a neighbouring flare, the dynamics and mixing in the cross-section should be even more dramatic, due to larger velocities and displacements. \par
This first study into the dynamics of the multi-stranded coronal loops has used a rather simple model. Future models should include a realistic atmosphere and footpoint driver, with included hydrodynamic evolution parallel to the magnetic field (thermal conduction, radiative losses, heating). With current computational power, and the limit on the size of the strands, this will be very challenging.\par
In conclusion, our simulations cast a strong doubt on the applicability of multi-stranded loop models, because such models are unstable when driven with transverse waves. We propose a reinterpretation of how the interior structure of a multi-stranded loop should be viewed. We should move away from the classical picture of a bundle of independent strands, because the perpendicular transport is much greater (due to KHI) than what is assumed to motivate the use of such multi-stranded models. Our models indicate that we should favour a loop structure of an ever-interacting and mixing, dynamic and inhomogeneous plasma, presenting structures in a  power-law-like, large range of scales.

\begin{acknowledgements} N.M. acknowledges the Fund for Scientific Research-Flanders (FWO-Vlaanderen). T.V.D. was supported by an Odysseus grant, the Belspo IAP P7/08 CHARM network and the GOA-2015-014 (KU Leuven). \end{acknowledgements}

\bibliographystyle{apj} 
\bibliography{../Biblio}{} 

\appendix
\section{Appendix} \label{Appendix}

\begin{figure}[h!]
    \centering
    	\includegraphics[width=0.5\textwidth]{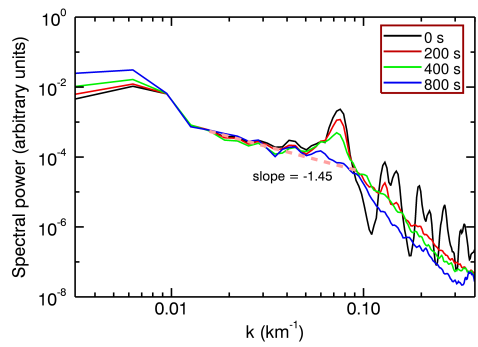}  
    	\caption{Power spectra for the density cross-section of the loop structure at $z=40$ Mm at four different times. Also shown is the linear fit to the central part of the power spectra at $t = 800$ s.}
    	\label{spectra2}
 \end{figure}

The growth rate of the Kelvin-Helmholtz instability, presented in Sec.~\ref{loop}, depends linearly on the amplitude of the transverse oscillation of the multi-stranded loop structure. Thus, the rate of mixing also depends on the amplitude. In this sense, even if this paper is not intended as a parametric study, it is insightful to see how a smaller amplitude driver than the one used in Sec.~\ref{driver} affects the internal structure of the loop. Eventually, we expect the smaller amplitude driver to cause the same degree of mixing, albeit after a longer time.
\begin{figure*}[h!]
    \centering
    	\includegraphics[width=0.9\textwidth]{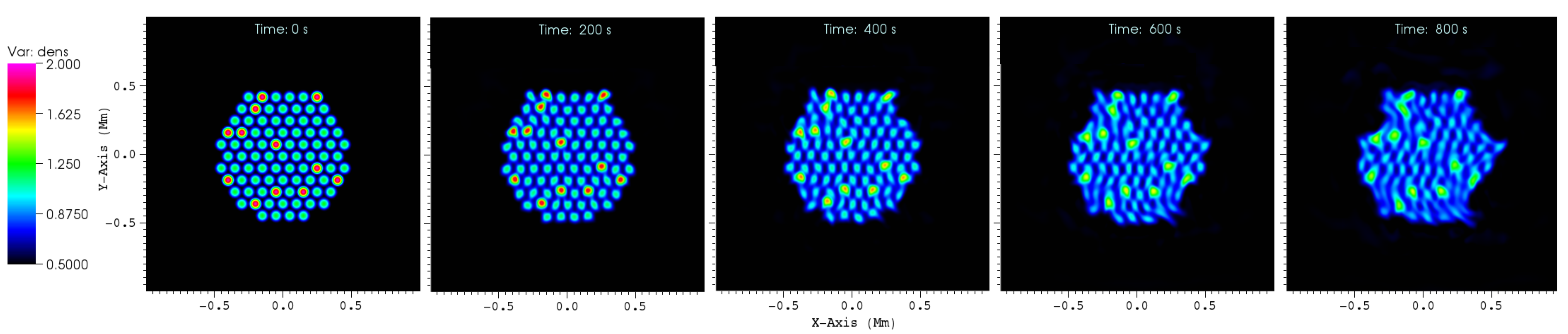}  
    	\caption{Plots of the loop cross-section density, at different times as indicated at the top of each panel. The slices are made at $z = 40$ Mm. The colorbar is common for the plots, in units of $10^{-12}$ kg m$^{-3}$.}
    	\label{img6}
 \end{figure*}
  We use the same driver as in Eq.~\ref{driver}, but choosing $A = 2.5$ km/s. The evolution of the cross-section can be viewed in Figure~\ref{img6}. The smaller amplitude of the driver leads to a slower mixing process in the cross-section, as expected. After $t = 800$ s, the mixing has a local appearance, i.e. affecting only the nearest neighbors of each strand. Still, we can spot some strands merging or breaking up, especially near the boundary of the loop structure, which is deformed.
  The evolution of the distribution of length-scales can be followed in Figure~\ref{spectra2}. This can be compared to Figure~\ref{spectra}. The slope of the central part of the power spectra is different, which can be explained by the less developed mixing, but the disappearance of a dominant length-scale (corresponding to the inter-strand distance) is present. Furthermore, we can observe the tendency of the central part of the slope to become steeper in time in both Figures. Thus, we expect the slope in the small-scale to tend towards the same value as in Figure~\ref{spectra} at later times, as the mixing progresses.

\end{document}